\documentclass[preprint,12pt]{elsarticle}

\usepackage{graphicx}
\usepackage{epsfig,array,a4}
\usepackage{latexsym}
\usepackage{amssymb}
\RequirePackage{amsthm,amsmath,natbib}

\newcommand{\bea}{\begin{eqnarray*}}
\newcommand{\eea}{\end{eqnarray*}}
\newcommand{\beq}{\begin{equation}}
\newcommand{\eeq}{\end{equation}}

\newcommand{\e}{\epsilon}
\newcommand{\s}{\sigma}

\newcommand{\Var}{\mbox{Var}}
\newcommand{\Cov}{\mbox{Cov}}

\def \R {\mathchoice {\hbox{I\kern -0.1667em R}}{\hbox{I\kern -0.1667em R}}
                      {\small{I\kern -0.1667em R}}{\small{I\kern -0.1667em R}}}

\begin{document}

\begin{frontmatter}



\title{A model selection approach to genome wide association studies}


\author[vie]{Florian Frommlet}
\author[vie]{Felix Ruhaltinger}
\author[wroc]{Piotr Twar\'og}
\author[wroc]{Ma{\l}gorzata Bogdan}

\address[vie]{Department of Statistics, University of Vienna, Austria}
\address[wroc]{Institute of
Mathematics and Computer Science, Wroc{\l}aw University of Technology, Poland}

\begin{abstract}
For the vast majority of genome wide association studies (GWAS)
published so far, statistical analysis was performed by testing
markers individually. In this article we present some elementary statistical considerations
which clearly show that in case of complex traits the approach based on multiple regression or generalized linear  models is preferable to multiple testing.  We  introduce a model
selection approach to GWAS based on modifications of Bayesian Information Criterion (BIC) and
develop some simple search strategies to deal
with the huge number of potential models. Comprehensive
simulations based on real SNP data confirm that model selection
has larger power than multiple testing to detect causal SNPs in
complex models. On the other hand multiple testing has substantial problems with proper ranking of causal SNPs and tends to
detect a certain number of false positive SNPs, which are not linked to any of the causal mutations. We show that this behavior is typical in GWAS for complex traits and  can be explained  by
an aggregated influence of many small random sample correlations between genotypes of a SNP under investigation and other causal SNPs. We believe that our findings at least partially explain problems with low power and nonreplicability of results  in many real data GWAS. 
Finally, we discuss the advantages of our model selection approach in
the context of  real data analysis, where we consider publicly available
gene expression data as traits for individuals from the HapMap project.

\end{abstract}


\end{frontmatter}



\section{Introduction}

Within the last five years genome wide association studies (GWAS) have become an important tool for genetic scientists. There exist several excellent reviews which elucidate the statistical intricacies involved, see e.g. \cite{Bal,HD,McC,ZKT}. The major goal of GWAS is to detect association between some trait (either quantitative or categorical) and some genetic markers. The most commonly used type of markers are single nucleotide polymorphisms (SNPs). Current SNP array technology allows to determine the state of up to one million  SNPs within a single experiment.

The huge number of markers leads to a multiple testing problem which has been extensively discussed in the literature (see the discussion on this topic in  \cite{ZKT}).
It is common practice in applied papers on GWAS to report single marker tests of SNPs. For a review on statistical tests for case control studies we refer again to \cite{ZKT}.  Recommended significance levels to control family wise error are as small as $\alpha = 5 \cdot 10^{-8}$   \cite{DG}, though occasionally larger significance levels like $\alpha = 10^{-6}$ are used. It is well understood that due to positive correlations between markers a simple Bonferroni correction is likely to be too conservative. Approaches  to deal with correlation between SNPs include permutation tests like in \cite{Strange3} or the use of Hidden Markov Models \cite{Wei}.

Most GWAS are performed as case control studies, though recently there has been growing interest in GWAS for quantitative traits (QT) \cite{Pot_b}.  Statistical analysis performed for QT tends to make use of regression models to correct for covariates like sex or age; however, each SNP is then tested individually at a significance level accounting for multiple testing (see for example \cite{Cho, Gan, Mei, New, Pot_a} etc.).

In the fairly related area of QTL mapping based on designed breeding experiments  the search over single markers was abandoned already quite a while ago in favor of multi marker models. In this context the problem of selection of significant markers is equivalent to the choice of the ``best'' multiple regression or generalized linear model.  This task is however rather difficult due to the large number of potential regressors.  Specifically, in \cite{BS}  it was noticed that classical model selection criteria like Akaike Information Criterion (AIC, \cite{AIC}), and even Bayesian Information Criterion 
(BIC, \cite{SCH}), tend to select too many markers. Addressing this problem \cite{BGD} introduced a modified version of BIC (mBIC), suited for the situation where a large number of markers is searched, but only relatively few markers are expected to be true signals. mBIC was motivated
in a Bayesian setting, using informative priors on the model dimension, which prefer rather small models. In \cite{BF} and \cite{BZG} it was observed that   mBIC penalty is closely related to the multiple regression version of the Bonferroni correction for multiple testing.
 In a series of papers \cite{Bai, BFBB, BF, BZG, EBC, ZBBF}  based on simulation studies good properties of mBIC were documented. Recently some asymptotic optimality properties of mBIC have been shown  \cite{BCG, FBC}.

In this article we will adopt a model selection approach for GWAS, which will be introduced in Section \ref{Sec:Meth}.
For the ease of presentation we will restrict our discussion mainly to linear regression models, though qualitatively similar results will hold for generalized linear models.  In Section \ref{Sec:Lin} we present basic statistical considerations, which clearly demonstrate the advantage of using  multiple regression over the search over single markers. In this section we specifically stress the lower power of single marker tests. This results from an inflated residual sum of squares, which incorporates the influence of all causal genes that are not included in the model. In Section \ref{Sec:mBIC} we introduce and motivate our particular choice of  model selection criteria for the high dimensional multiple regression we are dealing with. Apart from mBIC we  consider a second modification of BIC, mBIC2,  which  according to \cite{FBC} is closely related to the multiple regression version of the Benjamini-Hochberg procedure \cite{BH} for multiple testing. According to \cite{FBC} mBIC2  has  asymptotic optimality properties in a wider range of sparsity parameters than mBIC.
The greatest challenge in applying model selection to GWAS is the huge search space of potential models. In Section \ref{Sec:Search} we describe some model search strategies which are particularly suited to the situation where we have a huge number of markers but expect only a small number of them to be strongly associated with the trait.

In Section \ref{Sec:Sim} we  perform a  simulation study based on actual SNP data.
Apart from the expected result that model selection strategies outperform multiple testing procedures the simulation study provides some rather surprising insights. In particular we will see that for multiple testing procedures rather small random correlations  between causal SNPs have drastic influence on the order of p-values. This results in a low power to detect some important causal mutations, as well as in a large number of spurious detections. Thus, our results show that single marker tests can lead to many erroneous results, which makes the use of these procedures in the context of GWAS rather questionable.

Finally in Section \ref{Sec:Real} we reanalyze publicly available gene expression data  \cite{Strange1, Strange2, Strange3} as quantitative traits for the individuals genotyped in the HapMap project \cite{Hap}.  One aim of \cite{Strange3} was to detect   association with gene expression levels of SNPs lying outside the region of the considered gene (trans regulatory SNPs).
 44 genes with trans regulatory SNPs were reported when pooling over all HapMap populations. Using our model selection approach  we are able to increase the number of detected trans regulatory regions in several cases.

\section{Methods}\label{Sec:Meth}

We first want to motivate why we advocate a model selection approach. The discussion in this article is in the context of linear regression models, which allows to focus on the principal ideas involved. We expect that similar conclusions as presented here will also hold for generalized linear models, and in particular for logistic regression, which might be applied in case control studies.

\subsection{Linear regression models}\label{Sec:Lin}

 Let $y_i, i \in \{1,\dots,n\}$, denote measurements of a quantitative trait in $n$ individuals. Assume there are $p$ SNPs where $p \gg n$, but only $k \ll n$ of them have an effect on $y$. Let $\mathbf{j}^{\star} = (j^{\star}_{l_1}, \ldots, j^{\star}_{l_k})$, with $1\leq j^{\star}_{l_1}< \ldots<j^{\star}_{l_k}\leq p$, denote the ordered set of indexes of causal SNPs. We denote the genotype of person $i$ and SNP $j$  as $x_{ij} \in \{-1, 0, 1\}$ and assume that the additive model
\beq\label{AddModel}
{\cal M}_{\mathbf{j}^{\star}}: \quad y_i = \beta_0 + \sum_{l=1}^k \beta_{j_l^*} x_{i j_l^*} + \e_i
\eeq
holds.  For the sake of simplicity  we assume that the error terms are i.i.d. normal random variables   $\e_i \sim {\cal N}(0,\sigma^2)$. In a more realistic scenario the  model can be easily extended by including  other covariates, like sex or age.

The model proposed in (\ref{AddModel}) is rather simple. However, complexity arises because the task is to find this model among $2^p$ possible models (we consider only models including the intercept). This is a gigantic number taking into account that in GWAS we are usually dealing with $p \approx 10^7$.
The null model which does not include any causal SNP will be denoted by ${\cal M}_0$.
All further additive models can be characterized by multi indices ${\cal M}_{\mathbf{j}}$, where $j$ is an ordered subset of elements of the set $\{1,\ldots,p\}$.  Generically we will write $q = q_{\mathbf{j}}$ for the number of markers of a model.   For the correct model we have  $q_{\mathbf{j}^*} = k$.

Define for each model  ${\cal M}_{\mathbf{j}}$   the matrix $X^{\mathbf{j}} = (\mathbf{1}, x_{j_1}, \dots  x_{j_q})$, where $\mathbf{1} = (1,\dots,1)'$. Then we have in vector notation
\beq\label{AddModel_general}
{\cal M}_{\mathbf{j}}: \quad  y = X^{\mathbf{j}} \beta_{\mathbf{j}} + \e^{\;\! \mathbf{j}}
\eeq
 where $\beta_{\mathbf{j}} = (\beta_0,\beta_{j_1},\dots, \beta_{j_q})'$.
 Given the large number of markers it is understandable that it is common practice  to perform only single marker analysis. This means that only models of the form
\beq\label{SM_Model}
{\cal SM}_{j}: \quad y_i = \beta_0 +  \beta_j x_{ij} + \e_i^{(j)}
\eeq
are considered.

\begin{figure}[tbh]
   \centering
   \includegraphics[width=0.2\linewidth,  bb = 200 200 350 550]{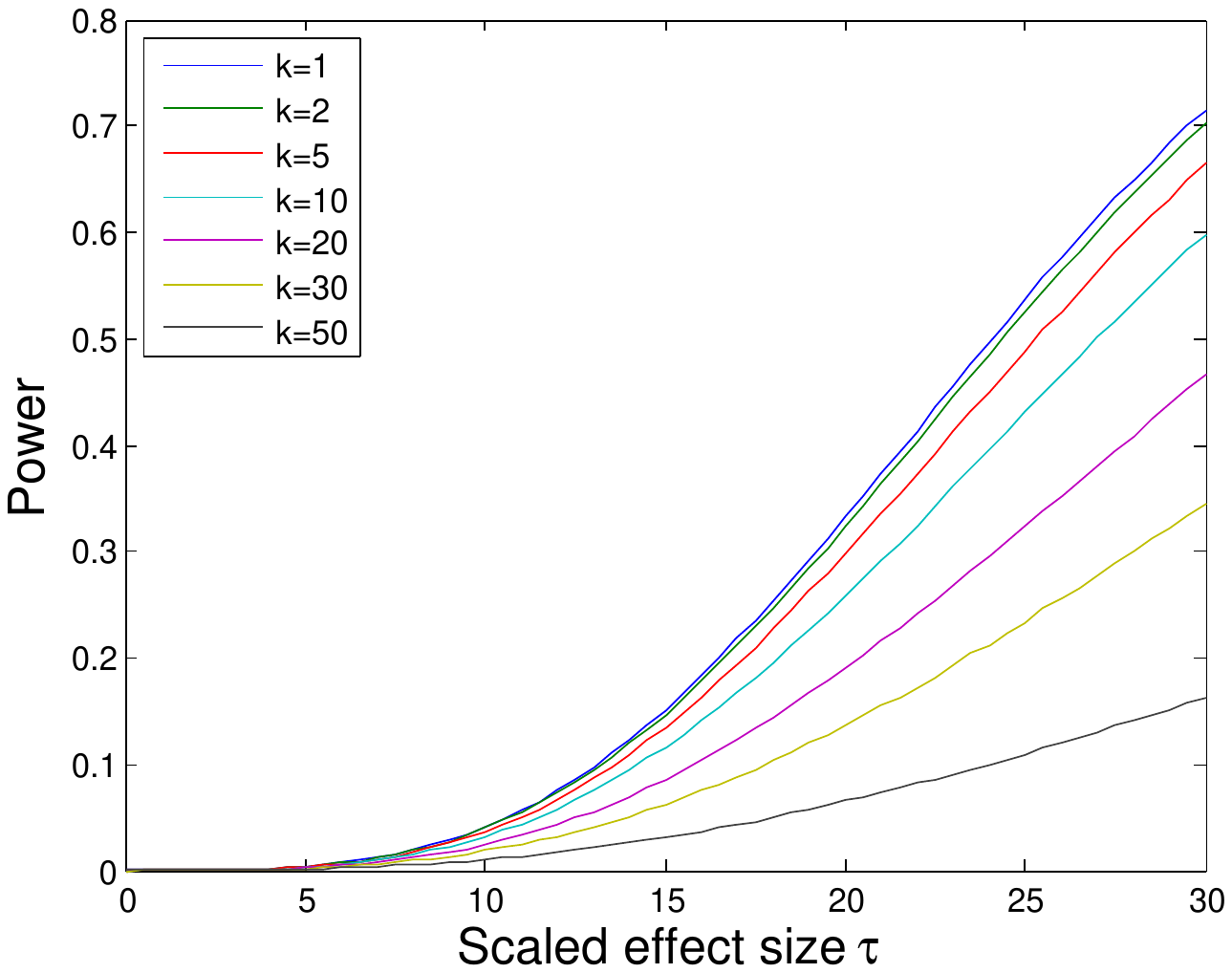}
\caption{Power to find a causal SNP with single marker tests when model ${\cal M}_{\mathbf j^*}$ with $k$ effects is correct.  In case of $k=1$ one is testing for the correct model.  }  \label{Fig:power}
\end{figure}

Elementary statistics tells us what happens when we perform an F-test for some model ${\cal M}_{\mathbf{j}}$ based on least squares regression.  Let $P_{\mathbf{j}} = (X^{\mathbf{j}})'[X^{\mathbf{j}} (X^{\mathbf{j}})']^{-1}X^{\mathbf{j}}$ denote the usual hat-operator for a general model ${\cal M}_{\mathbf{j}}$. Then
$RSS_{\mathbf{j}} := y'(I - P_{\mathbf{j}})y $ is the residual sum of squares and
$MSS_{\mathbf{j}} := y'( P_{\mathbf{j}} - \frac{1}{n}E)y$ is the model sum of squares for  ${\cal M}_{\mathbf{j}}$.
We always denote by $I$ the identity matrix and by $E = \mathbf{1} \mathbf{1}'$ the all one matrix of suitable dimension (here they are $n \times n$).   The usual F-test statistic  for the null hypothesis that none of the variables in the model   ${\cal M}_{\mathbf{j}}$ has an influence on $Y$  is given by
$$
F_{\mathbf{j}} = \frac{(n-q_\mathbf{j}-1)MSS_{\mathbf{j}}}{q_\mathbf{j} RSS_{\mathbf{j}}} \; .
$$
 {When the model ${\cal M}_{\mathbf{j}}$  includes all causal SNPs then the statistics $F_{\mathbf{j}}$} has a noncentral F-distribution and power calculations for different effect sizes  are rather straight forward (compare results for  $j= 1$ in Figure \ref{Fig:power}). However, we are often facing a different situation when model ${\cal M}_{\mathbf{j}^*}$ holds, but we are performing an $F$-test for a smaller model ${\cal M}_{\mathbf{j}}$, which might not include some of the causal SNPs.
Then
$$\e_i^{\mathbf{j}}= \quad y_i - \beta_0 - \sum_{l=1}^{q_j} \beta_{j_l} x_{i j_l} \sim  {\cal N}(\sum\limits_{l \in {\mathbf{j}}^* \backslash \{\mathbf{j}\}} \beta_l x_{il},\sigma^2)$$
 and  according to the generalization of Cochran's theorem for the noncentral case \cite{Mad} the model sum of squares and the residual sum of squares are independent with distributions
\begin{eqnarray} \label{noncentral_chi2}
\frac{MSS_\mathbf{j}}{\sigma^2} & \sim & \chi^2(q_\mathbf{j}, \frac{1}{\sigma^2} \beta_{{\mathbf{j}}^*}' (X^{{\mathbf{j}}^*})' (P_\mathbf{j} - \frac{1}{n} E) X^{\mathbf{j}^*} \beta{_{\mathbf{j}}^*} ) \; , \\
\frac{RSS_\mathbf{j}}{\sigma^2} & \sim & \chi^2(n-q_\mathbf{j}-1, \frac{1}{\sigma^2} \beta_{{\mathbf{j}^*}}' (X^{\mathbf{j}^*})' (I - P_\mathbf{j}) X^{{\mathbf{j}}^*} \beta_{{\mathbf{j}}^*} ) \; . \label{noncentral_chi2_b}
\end{eqnarray}
Here $\chi^2(u,v)$ denotes a noncentral chi-square distribution, with the number of degrees of freedom equal to $u$ and a noncentrality parameter $v$.
Thus the test statistic $F_{\mathbf{j}}$ is essentially the ratio of two independent noncentral $\chi^2$-distributed random variables. If the size of the true model $q_{\mathbf{j}^*}$ is much larger than the size $q_{\mathbf{j}}$ of the model under consideration, then the residual sum of squares $RSS_\mathbf{j}$ will have a considerably large noncentrality parameter incorporating effects which have not entered the model, and the power of the according $F$-test will be comparably small.

This effect will be most pronounced in case of simple regression models (\ref{SM_Model}).
To fix ideas consider for a moment the orthogonality assumption  $(X^{{\mathbf{j}}^*})' X^{{\mathbf{j}}^*} = n I$. Then the noncentrality parameters corresponding to $MSS_j$  and $RSS_j$ respectively become
\beq \label{nc_orthogonal}
\nu_{M,j} = \frac{n \beta_j^2}{\sigma^2} \quad \mbox{and} \quad
\nu_{R,j} =  \sum\limits_{l \in {\mathbf{j}}^* \backslash \{j\}} \frac{ n \beta_l^2}{\sigma^2}
\eeq
 In Figure \ref{Fig:power} power calculations are shown for this simplified situation with $n = 2000$ and $\alpha = 10^{-6}$. The squared scaled effect sizes $\tau = \frac{ n \beta_l^2}{\sigma^2}$ are equal for all $k$ effects. Power was obtained by sampling from the two noncentral $\chi^2$-distributions of (\ref{noncentral_chi2}) and (\ref{noncentral_chi2_b}).
If there is only a small number of causal SNPs the loss of power by testing for individual markers is not dramatic. However already for $k=10$ the loss becomes recognizable, and for $k=30$ one is actually losing more than 50\% of power in the range of effect sizes we considered.

Now in GWAS one can certainly not expect that all causal SNPs have the same effect size, and their genotypes will also not be orthogonal. However, GWAS are performed to understand the genetics of complex traits, which per definition are influenced by more than one factor.  Therefore our considerations concerning loss of power by single marker analysis will apply. For a single effects model ${\cal SM}_j$ one obtains
$$
(I - P_j) X^{\mathbf{j}^*} \beta_{{\mathbf{j}}^*} =
\sum\limits_{l \in {\mathbf{j}}^* \backslash \{j\}} \beta_l
\left(
(x_l -  \bar x_l) - \frac{\Cov(x_j, x_l)}{\Var(x_j)}
(x_j -  \bar x_j)
\right)
 $$
 where $\bar x_j$ is the sample mean and $\Var(x_j)$ and $\Cov(x_j, x_l)$ are the sample variance and covariance respectively.  The
 noncentrality parameters for single marker tests  have the form
\beq \label{nc_nonorthogonal_a}
\nu_{M,j} = \frac{\left(\sum_{l=1}^k \beta_l \Cov(x_j, x_l)\right)^2}{\sigma^2 \Var(x_j)}
\eeq
and
\beq \label{nc_nonorthogonal_b}
\nu_{R,j} =   \sum\limits_{l \in {\mathbf{j}}^* \backslash \{j\}} \sum\limits_{r \in {\mathbf{j}}^* \backslash \{j\}}
 \frac{ \beta_l\beta_r}{\sigma^2} \left(\Cov(x_l,x_r) - \frac{\Cov(x_l,x_j)\Cov(x_r,x_j)}{\Var(x_j)}
 \right)   \; .
\eeq

Compared to the case of orthogonality in (\ref{nc_orthogonal}) things are slightly more complicated due to correlation effects, which might have a strong influence on the noncentrality parameters for $RSS_\mathbf{j}$ and $MSS_\mathbf{j}$. In section \ref{Sec:Sim} we will show that the joint influence of many small random correlations between causal SNPs has a very strong effect on the noncentrality parameters $\nu_{M_j}$. This results in substantial problems with ranking of p-values and leads both to a low power of detection of some of the causal SNPs as well as to a relatively large number of false positives. Also, the general problem of loss of power when testing for single markers under a complex true model remains the same. We thus feel justified to claim that a model selection approach is the favorable alternative to single marker analysis.

\subsection{Modifications of BIC}\label{Sec:mBIC}

Assume that a family of models  ${\cal M}_{\mathbf{j}}$ has  parameters  $\theta_{\mathbf{j}}$ and corresponding likelihood functions  $L_{\mathbf{j}}(\theta_{\mathbf{j}})$. Denote by $\hat \theta_{\mathbf{j}}$ the maximum likelihood estimates of $\theta_{\mathbf{j}}$.
 Many statistical model selection criteria, like for example AIC or BIC,  suggest to select that model which maximizes a penalized likelihood function of the form
\beq
\log L_{\mathbf{j}}(\hat \theta_{\mathbf{j}}) - \eta\ q_{\mathbf{j}} \;.
\eeq
For AIC and BIC the penalty parameter  $\eta$ takes the form  1 and $\frac{1}{2} \log n$ respectively.

For linear regression under the assumption of normal error terms $\e^{\;\!\mathbf{j}} \sim{\cal N}(0, \sigma^2 I)$ the likelihood function of each model  ${\cal M}_{\mathbf{j}}$ in (\ref{AddModel_general}) is given by
$$
L_{\mathbf{j}}(y|\beta_{\mathbf{j}},\sigma) = \frac{1}{(\sqrt{2 \pi} \sigma)^n}
\exp\left( - \frac{(y - X^{\mathbf{j}} \beta_{\mathbf{j}})'(y - X^{\mathbf{j}} \beta_{\mathbf{j}})}{2 \sigma^2} \right) \; .
$$
The maximum likelihood estimator of
$\beta_{\mathbf{j}}$ then coincides with the least squares regression estimator $\hat \beta_{\mathbf{j}}$ and thus
$$
L_{\mathbf{j}}(y|\hat \beta_{\mathbf{j}},\sigma) = \frac{1}{(\sqrt{2 \pi} \sigma)^n}
\exp\left( - \frac{RSS_{\mathbf{j}}} {2 \sigma^2} \right) \; .
$$
For fixed $\sigma$ BIC is then equivalent to minimize
\beq \label{BIC_s_fixed}
  \frac{RSS_{\mathbf{j}}} {\sigma^2}  +  q_{\mathbf{j}}\log n  \;.
\eeq
For unknown $\sigma$ the ML-estimate $\hat \sigma^2 = \frac{RSS_{\mathbf{j}}}{n}$ leads to the criterion
\beq \label{BIC_s_unknown}
  n \log RSS_{\mathbf{j}}  +  q_{\mathbf{j}}\log n  \;.
\eeq

For sample size $n \geq 8$ the penalty parameter in AIC is smaller than in BIC and therefore in this case BIC tends to select more parsimonious models than AIC. Also, it is known that when $p$ is fixed and $n$ goes to infinity then  BIC is  consistent. Thus, when $n$ is large and  $p \ll n$, BIC usually selects the true model with large probability.
However, the situation is very different in case of $p > n$. 
As explained in detail in 
\cite{BZG} BIC is derived  with the underlying  prior assumption that all possible models ${\cal M}_{\mathbf{j}}$ are chosen with the same probability. This effectively results in using a Binomial prior $B(p,1/2)$ on the model dimension. Thus, BIC assigns a high prior probability to the class of models of size $p/2$, whereas small or very large dimensions are much less likely a priori.  Under sparsity, where the actual model has only a small number of regressors, this results in BIC choosing too many regressors.

As a remedy for this situation a modification of BIC was introduced in \cite{BGD}, which can be formulated  as
\beq \label{mBIC}
\mbox{mBIC:}  -2 \log L_{\mathbf{j}}(\hat \theta_{\mathbf{j}}) +  q_{\mathbf{j}}(\log n + 2 \log p + d) \;.
\eeq
This criterion was derived in a Bayesian setting assuming a prior probability of the model ${\cal M}_{\mathbf{j}}$ of the form
$$\pi({\mathbf{j}})=  \omega^{q_{\mathbf{j}}} (1 - \omega)^{p - q_{\mathbf{j}}} \; .$$
 In our context $\omega$ can be interpreted as the a priori expected proportion of causal SNPs, where all SNPs have independently from each other the same chance of being causal. This is a typical prior assumption in Bayesian model selection (see e.g. \cite{CGM}). Incorporating this prior distribution into BIC we easily obtain (\ref{mBIC}) with  $d = -2 \log(p \omega)$, i.e. minus two times the logarithm of the expected number of causal SNPs (for details of this derivation see \cite{BGD}).

In case of known $\sigma$ and under the assumption of orthogonal regressors mBIC has been shown to be closely related to the Bonferroni correction rule for multiple testing \cite{BZG}. In particular mBIC is controlling the family wise error. In \cite{FBC} it is shown that under certain sparsity conditions mBIC is consistent and has some optimality properties. Furthermore mBIC has been studied in the context of Generalized Linear Models \cite{Zak:2010} as well as Zero Inflated Generalized Poisson Regression \cite{EBC}.

Recently, in \cite{Chen1} a new modification of BIC, extended Bayesian Information Criterion (EBIC), was proposed. EBIC assigns a prior probability for the model dimension $q$, which is proportional to $\binom{p}{q}^{\kappa},$ for some
$\kappa \in [0,1].$
This results in the criterion
$$
\mbox{EBIC:}  -2 \log L_{\mathbf{j}}(\hat \theta_{\mathbf{j}}) +  q_{\mathbf{j}}\log n + 2\log\binom{p}{q}^{1-\kappa}\;.
$$
If $\kappa=1$ then EBIC coincides with BIC. The choice $\kappa=0$ corresponds to the uniform prior on the model dimension. In \cite{Chen1} some consistency properties of EBIC are proved under the assumptions that the maximal dimension searched by EBIC is fixed and larger then the true number of effects.  In \cite{Chen2} EBIC was further extended to Generalized Linear Models and in \cite{Chen3} it was successfully used for GWAS  with binary traits.

While EBIC turns out to work very well in many practical sparse cases, it has one undesirable property. When $q>\frac{p}{2}$ the last term of the penalty becomes a decreasing function of $q$ and encourages to pick the largest possible model. Therefore in this article we will consider a  slightly different criterion,
\beq \label{mBIC2}
 \mbox{mBIC2:} \quad -2 \log L_{\mathbf{j}}(\hat \theta_{\mathbf{j}}) +   q_{\mathbf{j}}(\log n + 2 \log p + d) - 2 \log (q_{\mathbf{j}}!) \;,
\eeq
which is asymptotically equivalent to EBIC  with $\kappa=0$ when the maximal allowable number of regressors, $Q$, is of the order $Q=o(p)$.

 mBIC2 was developed in \cite{FBC} as a model selection rule which in the context of multiple regression works similarly to the Benjamini-Hochberg correction for multiple testing.
In \cite{FBC} a thorough discussion is provided how this modification of mBIC relates to a similar criterion suggested by \cite{ABDJ} as well as to a modification of the risk inflation criterion RIC, proposed in \cite{GF}.
Due to the negative extra term  mBIC2 will potentially select larger models
than the original mBIC (\ref{mBIC}). In \cite{FBC} it is shown that mBIC2 has asymptotic optimality properties for a much larger range of sparsity levels than the original mBIC. We will compare the behavior of both criteria to select causal SNPs in the simulation study of Section  \ref{Sec:Sim}.

\subsection{Search algorithm} \label{Sec:Search}

An important question when applying a model selection approach to GWAS data is how to deal with the gigantic number of possible models.  Some  interesting search strategies for the best multiple regression model were recently proposed e.g in \cite{Chen3} and \cite{CC4}. However, these advanced model selection strategies are rather unfeasible for the large scale simulation studies. Therefore, for the purpose of our simulation study we developed our own search strategy, whose initial step relies on some modification of the popular forward selection.
Our method takes into account the fact that we are expecting a rather moderate number of causal SNPs (somewhere below 100) and turns out to be relatively accurate and fast enough to allow for a simulation study based on more than 300000 SNPs.  

In an initial step we perform single marker tests for all SNPs, a step we have to take in any case to be able to compare our model selection approach with the popular Bonferroni and Benjamini Hochberg (BH) procedures. For further analysis we only consider SNPs with an uncorrected p-value smaller than 0.15.

The second step consists of a simplified forward search strategy which we call multiple forward search. To this end
we start with computing the  original BIC (\ref{BIC_s_unknown}) for the single marker model  with lowest p-value.
Then we proceed iteratively by considering SNPs in ascending order of single marker p-values and
decide based on BIC to enhance the current model by a new SNP or not. This procedure is performed till we have considered all SNPs, or we have reached a maximum model size of 140 SNPs. For practical reasons we do not allow for larger models at this stage.

The initial multiple forward selection,  based on the uncorrected BIC, is expected to include a lot of false positives in the model, but hopefully also many causal SNPs. Its  principal advantage is that the actual search procedure,   based on modifications of BIC, is not starting from the null-hypothesis, but from a large model for which  the residual sum of squares will have been considerably reduced.
From here we start to perform backward selection and then stepwise selection based either on mBIC (\ref{mBIC}) or on mBIC2 (\ref{mBIC2}). This search strategy is designed to overcome the difficulties discussed in Section \ref{Sec:Lin}, when the actual model includes a large number of causal SNPs. It works well in the simulation study of Section \ref{Sec:Sim}, where more time consuming search procedures are out of question. In the real data analysis of Section \ref{Sec:Real}  the procedure will be amended with a final step, where all subsets of a specified set of SNPs are considered.

\section{Simulation study} \label{Sec:Sim}

Simulations are based on SNP data from the population reference sample POPRES \cite{Nel}.
The dataset used for simulations in this manuscript
was obtained from dbGaP through dbGaP accession number
phg000027.v2.p2 at \\www.ncbi.nlm.nih.gov/projects/gap/cgibin/study.cgi?study\_id=phs000145.v1.p1.  In particular we used data from the file Glaxo.txt, which contains a subsample of individuals studied in the article \cite{Lao}.  It comprises genotypes from 309788
SNPs for 649 individuals, which all have European ancestry and represent a relatively homogeneous population.


In this data set approximately 5\% of genotype values were missing.  To deal with these missing  values we adopted the following imputation strategy.
Suppose $x_{ij}$, the genotype of SNP $j$ for the $i$-th individual, is missing. We search for the 4 SNPs with strongest correlation to SNP $j$ fulfilling two conditions:
They are in a neighborhood of 500 SNPs upstream or downstream of SNP $j$ and their values for the $i$-th individual are not missing.
If we find individuals who have exactly the same values as the $i$-th individual on these 4 SNPs, then we predict the value of $x_{ij}$  as the most frequent value of  SNP $j$ among these individuals.
If we cannot find individuals fulfilling the above mentioned condition, then the most frequent value of the $j$th SNP among all individuals is imputed.

Since the main purpose of this experiment is to present some basic properties of different approaches to GWAS, both our simulation and search procedures treat the final set of obtained SNP genotypes as the ``correct'' one. Therefore, the imputation procedure has no influence on the final results.

Among the $p = 309788$ SNPs we have chosen $k = 40$ SNPs from autosomal chromosomes to be causal. These were selected deliberately in such a way that they are common and well distributed over all chromosomes.  The minimum allelic frequency for all causal SNPs was ranging between 0.3 and 0.5; variance of their genotype data was ranging between 0.42 and 0.53;  and correlations between all possible pairs of causal SNPs was  between  -0.12 and 0.1.  For the considered sample size this range of sample correlations corresponds well to the range of random sample correlations between independent SNPs.

We  simulated 1000 replicates from the additive model (\ref{AddModel}) where ${\mathbf j}^*$ indicates the 40 causal SNPs. Error terms were sampled from a standard normal distribution, i.e. $\s^2 = 1$. The 40 effect sizes were equally distributed between 0.27 and 0.66.
The overall heritability, defined as
\beq\label{Her_overall}
H^2 = \frac{\Var(X^{{\mathbf j}^*} \beta_{{\mathbf j}^*})}
{1 + \Var(X^{{\mathbf j}^*} \beta_{{\mathbf j}^*})} \;,
\eeq
is  equal to 0.81.  Heritability of an individual effect considered, defined as
\beq\label{Her_individual}
h_{j^*}^2 = \frac{\beta_{j^*}^2 \Var(x_{j^*})}
{1 + \Var(X^{{\mathbf j}^*} \beta_{{\mathbf j}^*})}\;,
\eeq
ranges between $ 0.006$  and  $ 0.037$.

We are aware of the fact that the overall heritability is unrealistically large, but we consider it  instructive to present the difficulties of the multiple testing procedures, which occur even in this simplified setting.  We believe that the phenomena discussed further in this paper, which result from a large number of causal SNPs, can play a role in explaining the problem of 'missing heritability' in GWAS \cite{Man}, a point which we extensively discuss in Section \ref{Sec:herit}.

Each simulated data set was analyzed using multiple testing procedures (Bonferroni and Benjamini Hochberg) as well as model selection approaches based on mBIC and mBIC2.
Bonferroni multiple testing correction was performed at family wise error rate  $\alpha = 0.05$, which corresponds to an  adjusted significance level of approximately $1.6 \cdot 10^{-7}$.
 Benjamini Hochberg  procedure was performed at the correponding FDR level $\alpha = 0.05$.
Model selection with mBIC was based on the constant $d = - 2 \log(4)$, which serves as a standard choice (see e.g. \cite{BZG}). Based on the calculations of \cite{BGD} we expect that for  $p$ and $n$ of this data set mBIC controls the family wise error under the total null approximately at a level  $\alpha = 0.02$.  We have also computed Bonferroni correction and BH at this smaller level, but given the observed lack of power of both BH and Bonferroni  these results are not presented.


In GWAS studies it frequently happens that not the causal SNP itself is detected as significant, but some SNP whose genotype is highly correlated to the causal SNP. Such a finding is not necessarily to be considered as a false positive, which leads to the question how to define true and false positives for correlated regressors. We adopt the following convention: Any detected SNP whose correlation to a causal SNP has absolute value larger than a given threshold is counted as a true positive, otherwise as a false positive. We initially used a threshold of $|R| = 0.9$, and based on simulation results decided to report alternatively also  results for a threshold $|R| = 0.7$. Here $|R|$ is the maximum absolute value over all correlations with causal SNPs.

For multiple testing procedures it frequently happens that two or more selected SNPs are correlated with a causal SNP. In case they are above the specified threshold they are all counted as just one true positive. False positive SNPs with identical genotype are only counted once. Based on these definition we estimate the power of detection for  each individual causal SNP, as well as the false discovery rate (FDR).

\begin{figure}[tb]
   \centering

\begin{minipage}{5cm} TP: $|R| > 0.9$ \\[3mm]  \centerline{\includegraphics[width = 2.5cm, bb = 200 200 350 550]{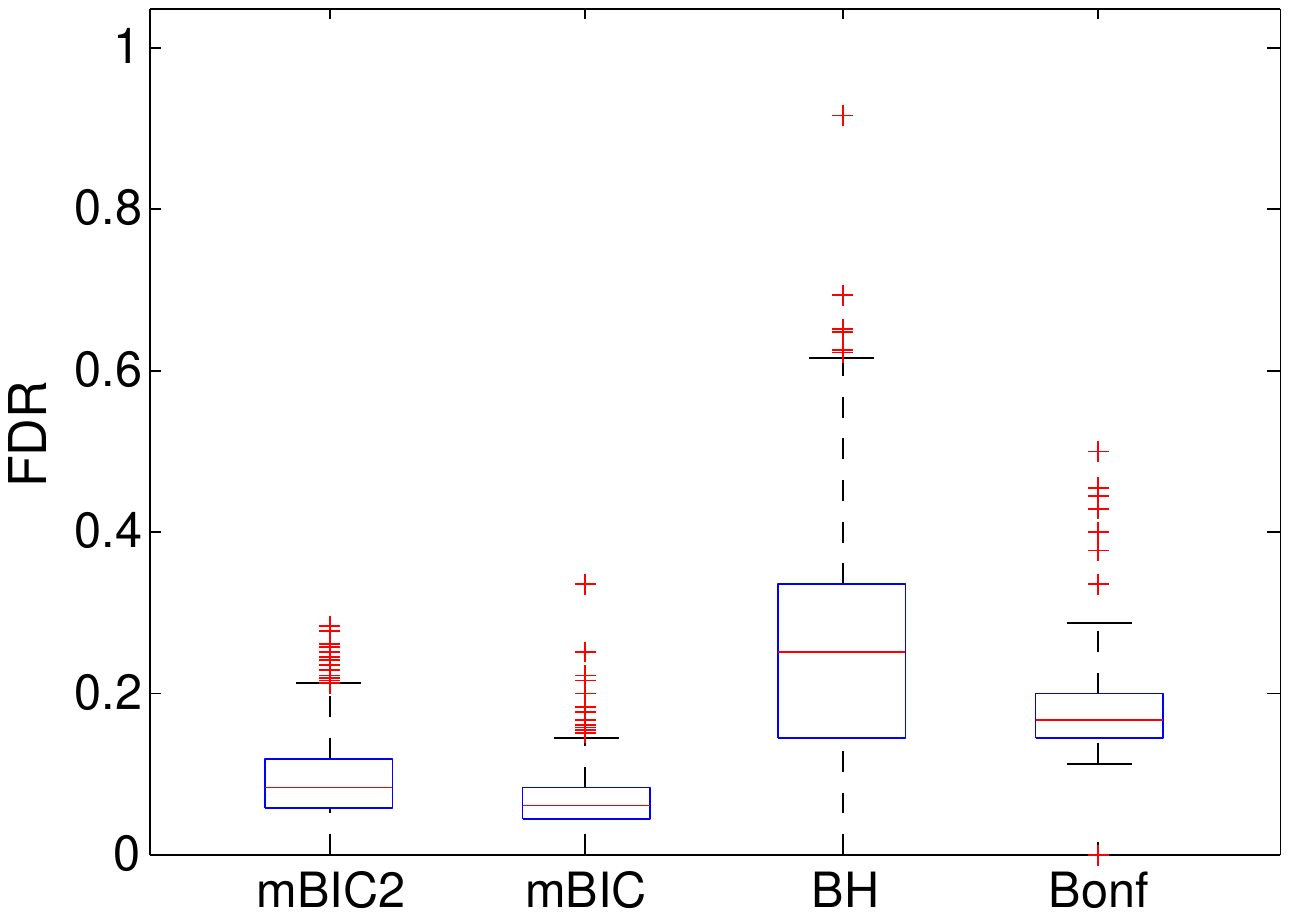}}
\end{minipage}
\hspace{2cm}
\begin{minipage}{5cm} TP: $|R| > 0.7$  \\[3mm]  \centerline{\includegraphics[width = 2.5cm, bb = 200 200 350 550]{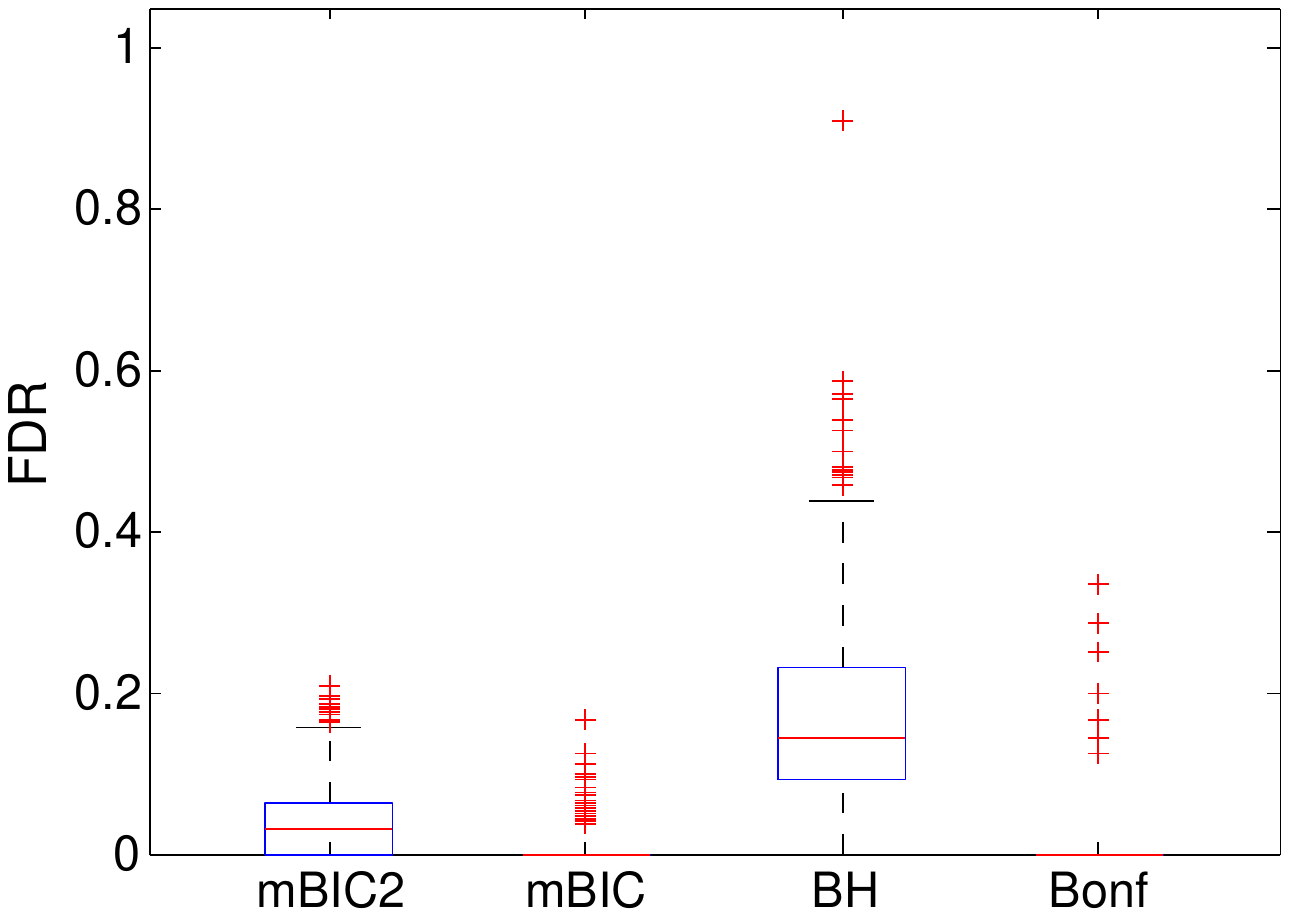}}
\end{minipage}
   \caption{Observed FDR of the 4 different selection procedures. A detected SNP was classified as a true positive when its maximal correlation to a causal SNP was larger than 0.9 in the first graph, and larger than 0.7 in the second graph. } \label{FDR}
\end{figure}

The graphs in Figure \ref{FDR} show the observed FDR values for all 1000 replicates based on the thresholds $|R| = 0.9$ and $|R| = 0.7$ respectively.   Both model selection procedures show much less variation in FDR than BH, which is a direct consequence of the fact that the model selection procedures have much larger power.  The false discovery rate of  BH at level 0.05 is apparently larger than  FDR of mBIC2 with the constant $d = -2 \log 4$, though in absolute terms the number of false positives is often smaller for BH.
The choice of the threshold for a ``false positive'' has a strong effect on  FDR. For all procedures FDR is significantly reduced  when using the more liberal criterion  $|R| = 0.7$. In particular the Bonferroni procedure detects in that case hardly any false positives.  The effect of the threshold on the number of false positives and on power is discussed in more detail in Section \ref{Sec:threshold}.

\begin{figure}[tb]
   \centering

\begin{minipage}{5cm} TP: $|R| > 0.9$  \\[3mm]  \centerline{\includegraphics[width = 2.5cm, bb = 200 200 350 550]{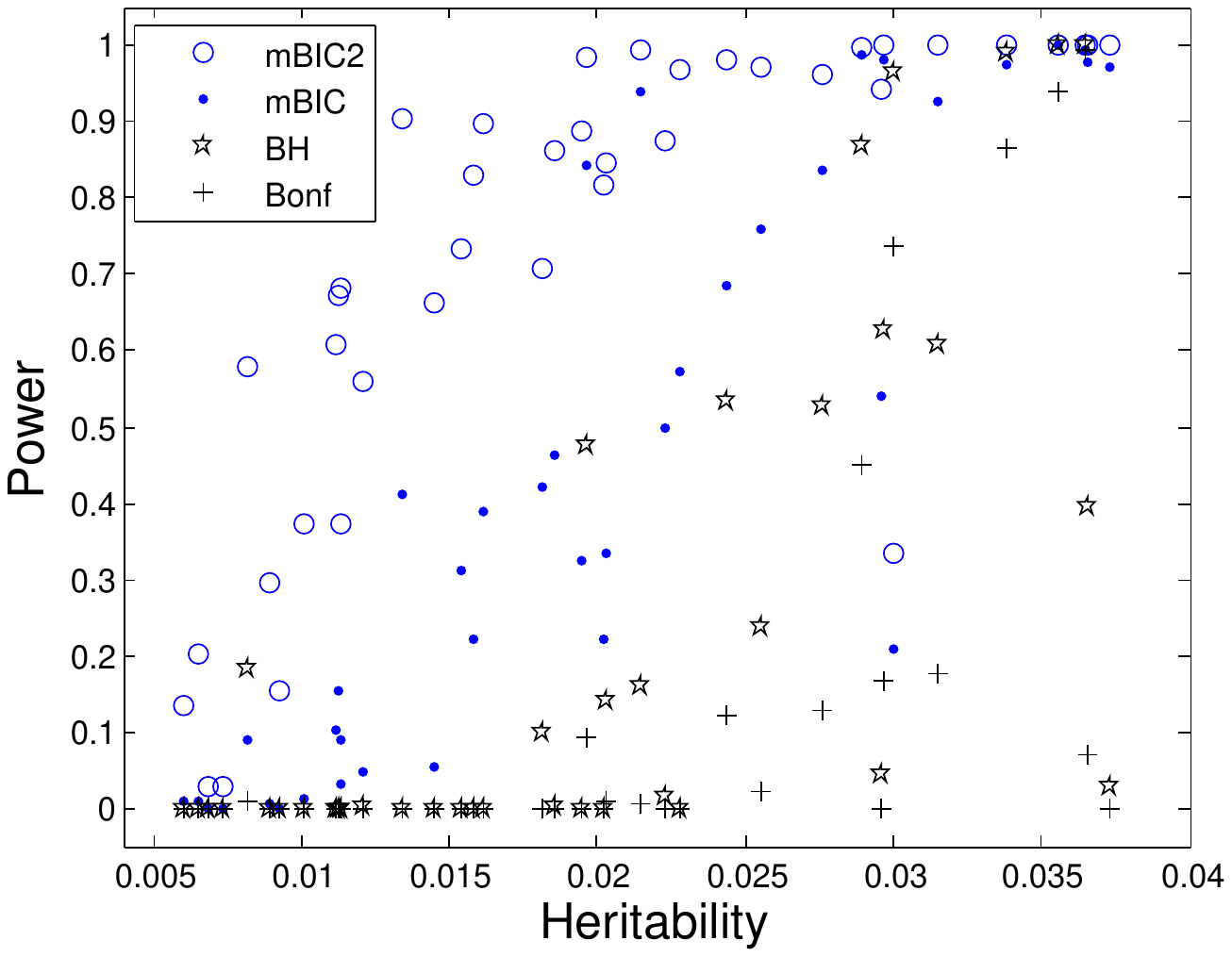}}
\end{minipage}
\hspace{2cm}
\begin{minipage}{5cm} TP: $|R| > 0.7$  \\[3mm] \centerline{\includegraphics[width = 2.5cm, bb = 200 200 350 550]{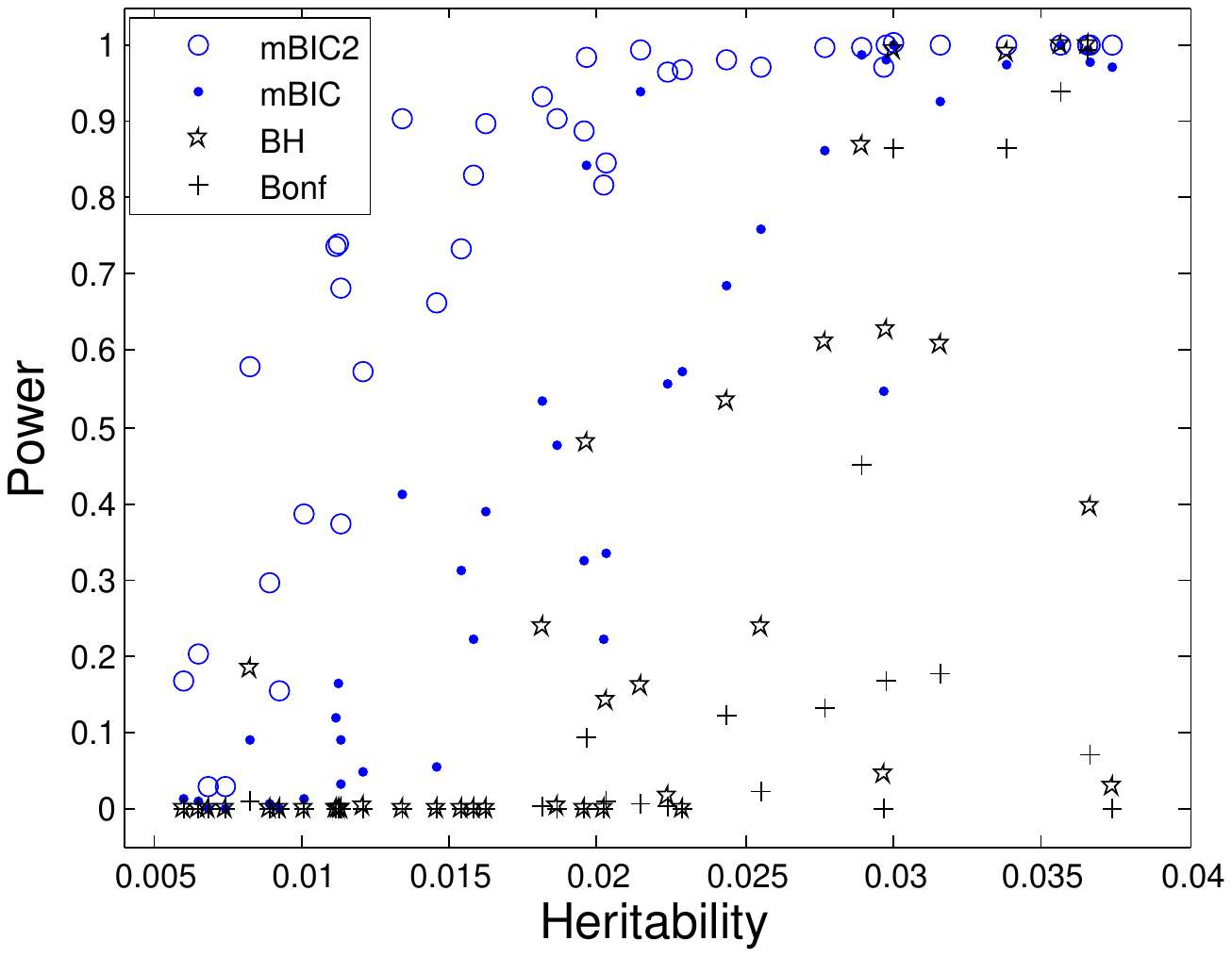}}
\end{minipage}
   \caption{Power of the 4 different selection procedures.}
   \label{Fig:Power1}
\end{figure}

The graphs in Figure \ref{Fig:Power1}  provide the estimated power to detect each of the 40 causal SNPs at the threshold levels $|R| = 0.9$ and $|R| = 0.7$  respectively. The $x$-axis shows the individual heritability (\ref{Her_individual}) of each SNP. It is evident that  mBIC2 has the largest power among the 4 different procedures. Also, as expected,  mBIC  has in most cases larger power than the two multiple testing procedures.

Most remarkable is the dependence of power on the individual heritability. As expected there is  a general trend that larger individual heritability yields larger power, but Figure \ref{Fig:Power1} shows that there is also a huge amount of irregularity. For mBIC2 in general the association between individual heritability and power is quite strong. Still, when using  threshold $|R| = 0.9$, there are several SNPs with relatively small power.  The most striking example is SNP A-1912140, with a large heritability of 0.03 and power of approximately only 33\%. This specific case will be explained in detail in Section \ref{Sec:threshold}.

For the  multiple testing procedures the behavior is much more erratic. The order of SNPs with respect to power is  entirely different from the order in terms of individual heritability. For example, the SNP with the largest heritability, easily detected by mBIC2 and mBIC, is completely missed by the procedures based on individual tests. On the other hand, some  substantially ``weaker'' SNPs, are detected with a power exceeding 50\%.   The key to understand this phenomenon is the influence of correlation between causal SNPs on the noncentrality parameter of the model sum of squares, as discussed in Section \ref{Sec:herit}. We will see that this has not only a negative effect on detecting causal SNPs, but it also gives rise to numerous detections of false positive SNPs which have no relation at all to the quantitative trait. This result we consider to be the most important outcome of this simulation study.

\subsection{Dependence of TP and FP on $|R|$ - thresholds  } \label{Sec:threshold}
Table \ref{Tab:FP} shows how often certain SNPs occur as false positives based on the threshold  $|R| = 0.9$ for mBIC2 and for BH.  The first significant finding is that  SNP A-2299101,  detected in 670 simulation runs, has a correlation of 0.8958 with causal SNP A-1912140. This explains the low power of 33\% to detect causal SNP A-1912140, and we conclude  that SNP A-2299101 is actually not really a false positive, but rather it is detected instead of SNP A-1912140. Thus the threshold  $|R| = 0.9$ to determine false positives is apparently  too strict.

\begin{table}
\caption{Thirty most frequent false positive SNPs (based on $|R| < 0.9$) under mBIC2 and under BH. First the SNP name, then the frequency in how many simulation runs the SNP was detected as a false positive, and finally the absolute correlation $|R|$ to the closest causal SNP.}  \label{Tab:FP}
\begin{center}
\begin{tabular}{rrl|rrl}
\multicolumn{3}{c|}{$\quad$ mBIC2} & \multicolumn{3}{c}{$\qquad$ BH} \\
SNP &  freq &  corr & SNP &  freq &  corr \\
\hline
 SNP\_A-2299101    & 670 & 0.8958  &  SNP\_A-2299101 & 990 & 0.8958 \\     
 SNP\_A-2034806    & 133 & 0.8416  &  SNP\_A-2181789 & 354 & 0.2628 \\     
 SNP\_A-2170607    &  92 & 0.7728  &  SNP\_A-1804206  & 136 & 0.7187 \\    
 SNP\_A-4270622    &  63 & 0.6255  &  SNP\_A-1839674  & 132 & 0.1137 \\    
 SNP\_A-2266375    &  56 & 0.8372  &  SNP\_A-1839540  & 128 & 0.5742 \\    
 SNP\_A-1790281    &  55 & 0.7683  &  SNP\_A-2251903 &  87 & 0.1116 \\     
 SNP\_A-2048646    &  47 & 0.8311  &  SNP\_A-2034806 &  61 & 0.8416 \\     
 SNP\_A-4201549    &  43 & 0.8351  &  SNP\_A-4236404 &  58 & 0.7087 \\     
 SNP\_A-1804206    &  35 & 0.7187  &  SNP\_A-1818215  &  56 & 0.1259 \\    
 SNP\_A-2101072    &  33 & 0.7659  &  SNP\_A-4201549 &  56 & 0.8351 \\     
 SNP\_A-2091172    &  31 & 0.7230  &  SNP\_A-2167803 &  52 & 0.0970 \\     
 SNP\_A-2299237    &  24 & 0.8954  &  SNP\_A-2048646 &  50 & 0.8311 \\     
 SNP\_A-4231385    &  23 & 0.8162  &  SNP\_A-1922491  &  50 & 0.7145 \\    
 SNP\_A-2267857    &  21 & 0.7871  &  SNP\_A-4291099 &  40 & 0.4563 \\     
 SNP\_A-2198243    &  18 & 0.8632  &  SNP\_A-1894129  &  39 & 0.5479 \\    
 SNP\_A-2293694    &  16 & 0.5097  &  SNP\_A-4217508 &  37 & 0.1007 \\     
 SNP\_A-2006296    &  15 & 0.7277  &  SNP\_A-1810532  &  34 & 0.1059 \\    
 SNP\_A-2119492    &  14 & 0.5207  &  SNP\_A-2241893 &  32 & 0.1373 \\     
 SNP\_A-1839540    &  12 & 0.5742  &  SNP\_A-2032742 &  32 & 0.1024 \\     
 SNP\_A-4266983    &  11 & 0.5207  &  SNP\_A-1804069  &  24 & 0.0896 \\    
 SNP\_A-1961183    &  10 & 0.8573  &  SNP\_A-1804341    & 19 & 0.0994 \\   
 SNP\_A-4241095    &   8 & 0.5421  &  SNP\_A-2120788   & 17 & 0.8978 \\    
 SNP\_A-1894737    &   8 & 0.6876  &  SNP\_A-2213672   & 16 & 0.1162 \\    
 SNP\_A-1784603    &   6 & 0.7480  &  SNP\_A-2171843   & 16 & 0.0997 \\    
 SNP\_A-2308622    &   6 & 0.7776  &  SNP\_A-2163734   & 14 & 0.1123 \\    
 SNP\_A-1957857    &   6 & 0.6638  &  SNP\_A-2294489   & 13 & 0.1109 \\    
 SNP\_A-4224627    &   5 & 0.1008  &  SNP\_A-4254512   & 13 & 0.0875 \\    
 SNP\_A-1965812    &   5 & 0.7632  &  SNP\_A-1865448    & 13 & 0.1115 \\   
 SNP\_A-1835435    &   5 & 0.1284  &  SNP\_A-2051237   & 12 & 0.3905 \\    
 SNP\_A-1976469    &   5 & 0.3733  &  SNP\_A-1816015    & 12 & 0.5199 \\   
\end{tabular}
\end{center}
\end{table}

Looking at the first column of Table \ref{Tab:FP} we observe that all SNPs detected by mBIC2 in more than 5 simulation runs have $|R|>0.5$.  Since none of the statistical approaches to GWAS can clearly distinguish  SNPs which are closely correlated, we believe that instead of reporting just one detected SNP, one should report also all SNPs which are strongly correlated to it.  The smaller the  suitable threshold the more SNPs one has to report, and to this end the value 0.5 appears to be fairly small. As a compromise we decided to consider $|R| = 0.7$ as a suitable threshold. With the exception of SNP A-4270622 all SNPs which are detected by mBIC2 in at least 18 simulation runs are then classified as true positives. 

On the other hand there is a relatively large number of SNPs which are detected by mBIC2 as false positives only once (1076), twice (55) or three times (12). These detections, which might be classified as actual false positives, are usually not correlated to any causal SNP.
Based on a model selection approach one would expect to detect some false positives of this nature, and their frequency is controlled according to the theory of mBIC2.

\subsection{Multiple testing procedures and heritability} \label{Sec:herit}

The second column of Table \ref{Tab:FP} shows the thirty most frequent false positive SNPs under BH. There are several SNPs which coincide with SNPs from the first column. Practically all other SNPs which have not been detected by mBIC2 as false positives have a striking characteristic: They are not strongly correlated to any causal SNP. The most prominent example is SNP A-2181789, which has been detected 354 times as a false positive, but is only correlated with $|R| = 0.2628$ to the closest causal SNP.

We will now provide the explanation for this seemingly implausible result, and will also explain the erratic behavior of power in terms of  individual heritability observed in Figure \ref{Fig:Power1}.
Remember that F-tests of single effect models involve non-centrality parameters (\ref{nc_nonorthogonal_a}) and (\ref{nc_nonorthogonal_b}). We can rewrite the square root of (\ref{nc_nonorthogonal_a}) as
$$
\sqrt{\nu_{M,j}} = \left|\frac{\beta_j} {\sigma} \sqrt{ \Var(x_j)} + \frac{ \sum_{l\neq j} \beta_l \Cov(x_j, x_l)}{\sigma \sqrt{\Var(x_j)}}\right|  \; .
$$
This shows that in case of orthogonal design matrix,  $\nu_{M,j}$ is proportional to the individual heritability. However, in the general case the non-centrality parameter is modified according to  $\sum_{l\neq j} \beta_l \Cov(x_j, x_l)$. This term can occasionally become  fairly large when there is a large number of true signals. Note that we designed our simulation study such that pairwise correlation between SNPs was small. In a statistical sense the genotypes of the causal SNPs can be thought of as being independent. Still, for some causal SNPs the effects of correlation just by chance accumulate significantly.

\begin{figure}[tb]
   \centering
\begin{minipage}{5cm} Power  \\[3mm]  \centerline{\includegraphics[width = 2.5cm, bb = 200 200 350 550]{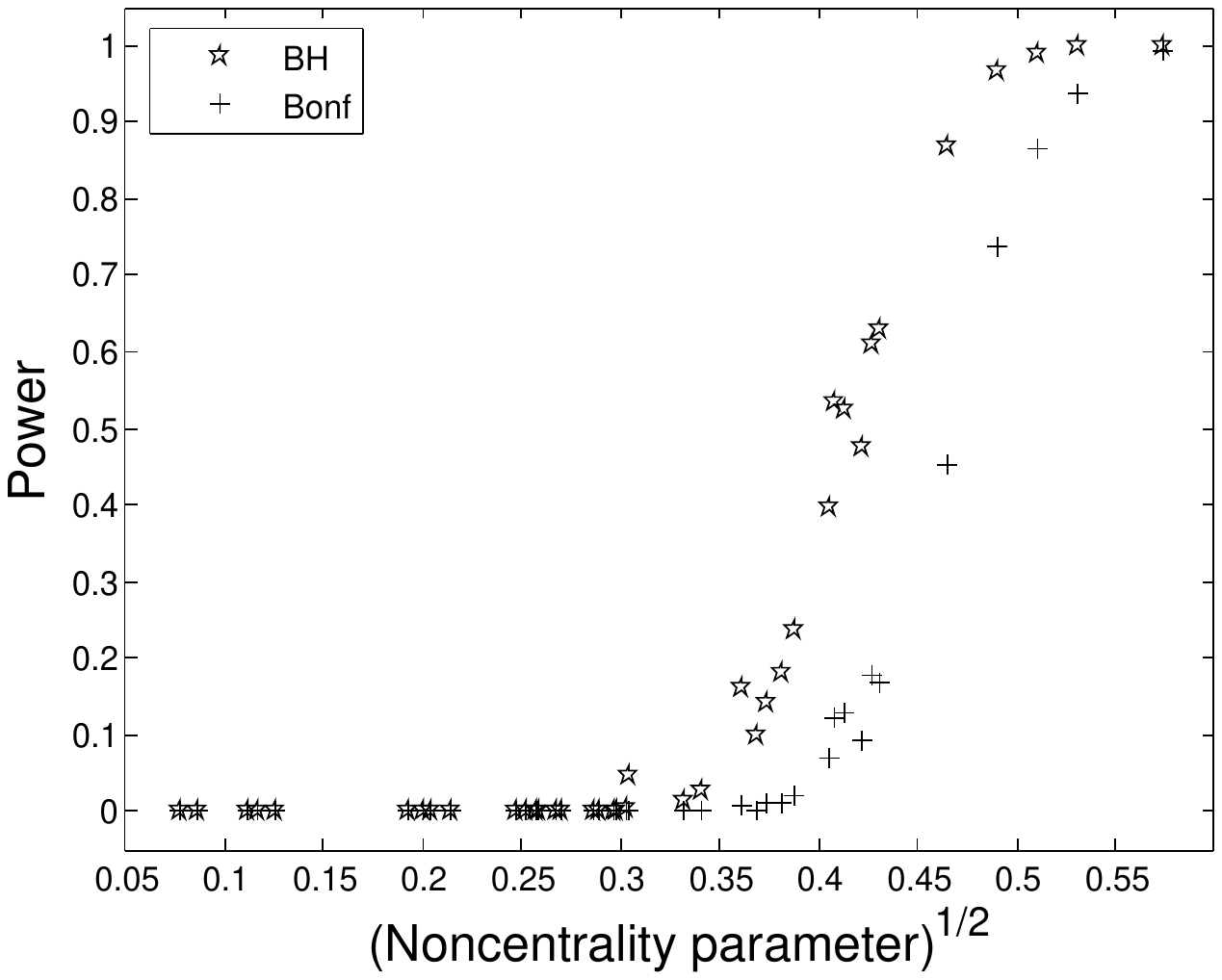}}
\end{minipage}
\hspace{2cm}
\begin{minipage}{5cm} NCP of FP \\[3mm] \centerline{\includegraphics[width = 2.5cm, bb = 200 200 350 550]{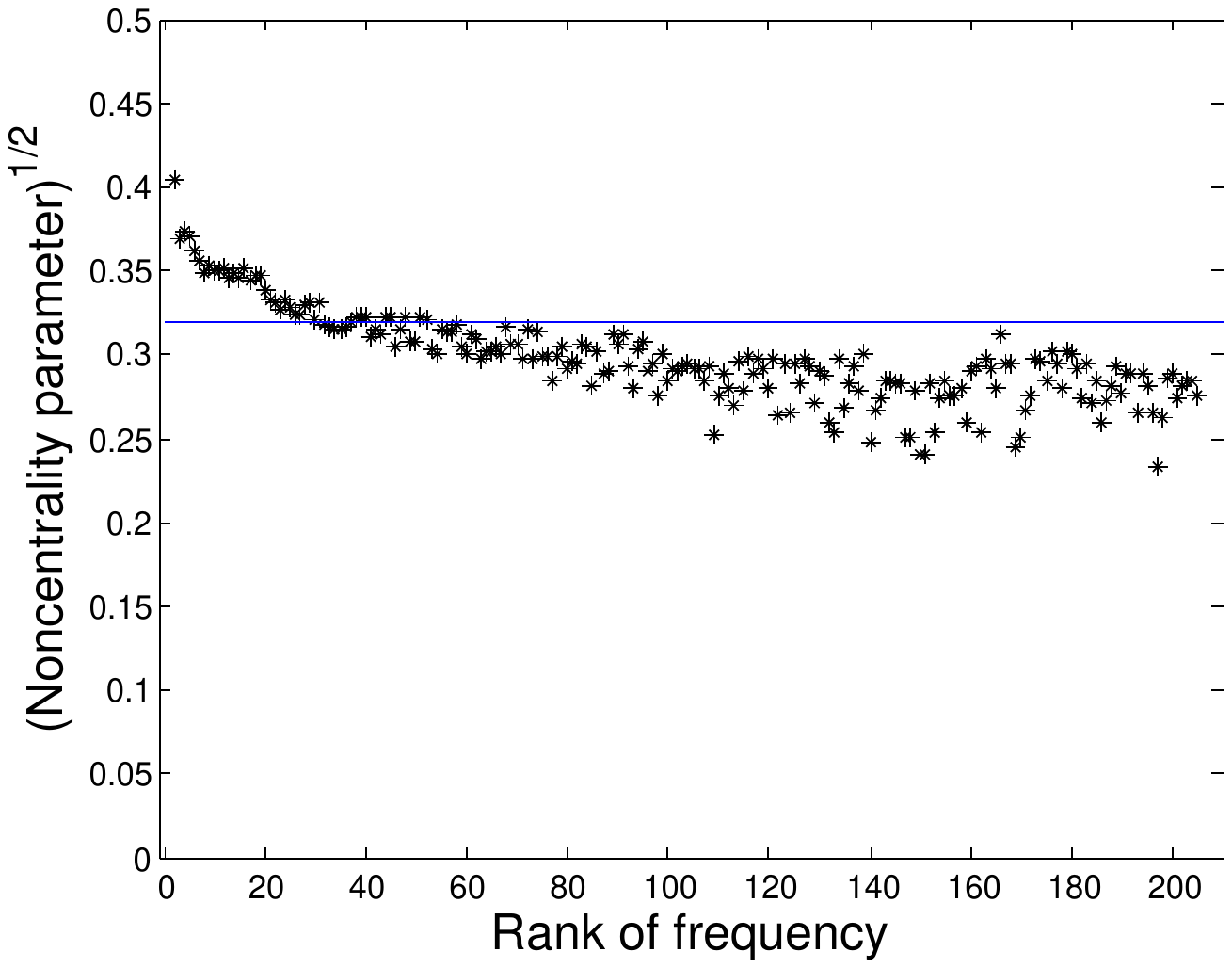}}
\end{minipage}
  \caption{First plot: Power of the multiple testing procedures. Instead of the individual heritability we plot against the square root of the noncentrality parameter.
Second plot: Square root of the noncentrality parameter of correlations for all false positives occurring under BH (at level $|R| = 0.9$). On the x-axis SNPs are ordered according to their frequency of detection in the 1000 simulation runs. The initial first stars from the left correspond to the SNPs listed in the second column of Table \ref{Tab:FP}.} \label{Fig:noncentrality}
\end{figure}

The first plot in Figure \ref{Fig:noncentrality} shows that small pairwise correlations with other causal SNPs  explain the erratic behavior of the multiple testing procedures. When we plot the power against the square root of the non-centrality parameter $\nu_{M,j}$ we observe the regular behavior of a sigmoid function. Clearly not the individual heritability but the weighted sum of correlations to all causal SNPs from (\ref{nc_nonorthogonal_a}) determines the power to detect an individual SNP.  This observation is crucial. It calls into question the practice of reporting detected SNPs according to the order of p-values from multiple testing procedures and claiming that SNPs with smallest p-values are the most important ones.  It might as well be the case that such signals are just catching the effect of many other causal SNPs which themselves are not detectable.  Also, since most of  the detected pairwise correlations between ``false'' SNPs and the trait result only from  random fluctuations of sample correlation coefficients between these SNPs and the causal ones, they are not replicable in different samples from the same population. Therefore, such ``false positives'' are useless also in terms of predicting the trait values.  

The second plot in Figure \ref{Fig:noncentrality} gives the answer to the question why some SNPs occur so frequently as false positives when they are not at all correlated with any causal SNP. It turns out that all false positive SNPs under BH have relatively large noncentrality parameter ($\sqrt{\nu_{M,j}} > 0.23$), and in particular those 30 SNPs listed in the second column of Table \ref{Tab:FP} all have  $\sqrt{\nu_{M,j}} > 0.32$. This is just the onset of the sigmoid function observed in the first plot of Figure \ref{Fig:noncentrality} where the power to detect causal SNPs no longer vanishes.

The conclusion from this analysis is the following. If we believe that a trait in a genome wide association study is influenced by a relatively large number of genes, then multiple testing procedures have a large chance of missing many of these genes. On the other hand there is a  large chance of detecting false positive SNPs which have nothing to do with functional regions. This appears to be quite a devastating r\'{e}sum\'{e} for the  performance of multiple testing procedures in GWAS with complex traits.

\section{Real data analysis} \label{Sec:Real}

In a series of papers Stranger et al. \cite{Strange1, Strange2, Strange3} have analyzed the association between SNPs from the HapMap project \cite{Hap} and gene expression data. In particular in \cite{Strange3} they considered  270 individuals from four populations, namely 30 Caucasian trios of northern and western European background (CEU), 30 Yoruba trios from Ibaden, Nigeria (YRI), 45 unrelated individuals from Beijing, China (CHB) and 45 unrelated individuals from Tokyo, Japan (JPT). The major objective of \cite{Strange3} was to find so called \emph{cis} associations and \emph{trans} associations between SNPs and gene-expression, where the cis region for SNPs was defined to be 1-Mb upstream or downstream of the expression probe midpoint.

For statistical analysis  \cite{Strange3} used a permutation test approach for test statistics obtained with simple linear regression models only considering additive effects. Excluding the 60 children from the CEU and YRI trios left 210 unrelated individuals. Analysis was performed considering the four populations separately, as well as combining data from individuals of certain populations.  Pooling over all four populations provides the most powerful approach, and we will thus restrict our analysis to this situation.
 To deal with population structure \cite{Strange3} used a procedure based on conditional permutations, which was originally described in  \cite{Kor}.

\begin{table}
\caption{Summary statistics for the 44 genes reported in \cite{Strange3}. Col. 2: Number of tag SNPs representing  detections by \cite{Strange3} for cis and trans association \{original number in curly brackets\}. Col. 3 and  4: Number of SNPs detected by mBIC2 as well as number of matches with and without taking into account population structure [number of cis SNPs within box]. Col. 3 also shows p-values of F-Test for dummy variables. Col. 5: Categories of genes as described in the main text.   }\label{Tab:Results_Real}
\vspace{-5mm}
\begin{footnotesize}
\begin{center}
\begin{tabular}{r|rr|rrl|rr|l}
& \multicolumn{2}{c|}{Stranger}  & \multicolumn{3}{c|}{with Dummy} & \multicolumn{2}{c|}{no Dummy} &\\
Gene            & Cis               &  Trans        &   SNPs &  Match  & pval   & SNPs    &     Match &  Cat   \\
GI\_14277699-S  &                   &     1         &  1 & 1 & 1,3E-16    &        5 &              1  & A         \\
GI\_15718725-S  &                   &     1         &  1 & 1 & 3,3E-08    &        3 &              1  & A         \\
GI\_21536317-S  &                   &     1         &  2 & 1  & 1,9E-05   &        6 &                 & A          \\
GI\_22749298-S  &                   &   \{3\}     1 &  1 & 1  & 4,6E-05   &        2 &              1  & A          \\
GI\_25952101-I  &                   &     1         &  1 & 1  & 1,6E-08   &        2 &              1  & A         \\
GI\_34147704-S  &                   &     \{3\} 1   &  1 & 1  & 5,6E-13   &        2 &              1  & A         \\
GI\_37545699-S  &                   &     1         &  1 & 1   & 2,3E-09   &       3 &                 & A          \\
GI\_37552052-S  &                   &     1         &  1 & 1  & 4,4E-20   &        3 &              1  & A         \\
GI\_39841070-S  &                   &     1         &  1 & 1  & 1,4E-14   &        3 &              1  & A         \\
GI\_41147791-S  &                   &     \{3\} 1   &  \fbox{1} 4 & 1  & 1,4E-23&  3 &              1  & A         \\
GI\_42655578-S  &                   &     1         &  1 & 1  & 2,8E-17   &        2 &                 & A         \\
GI\_42656964-S  &                   &     1         &  1 & 1  & 0,0023    &        2 &              1  & A         \\
GI\_42659691-S  &                   &     1         &  1 & 1  & 2,8E-08   &        3 &              1  & A         \\
GI\_42662536-S  &                   &    \{15\} 1   &  2 & 1  & 1,2E-09   &        1 &              1  & A         \\
hmm26651-S      &                   &     \{3\} 2   &  5 & 2  & 2,3E-06   &        8 &              1  & A         \\
hmm32074-S      &                   &     1         &  11 & 1  & 1,4E-42   &       9 &              1  & A         \\
hmm32535-S      &                   &     1         &  1 & 1  & 5,2E-22   &        5 &              1  & A         \\
Hs.292310-S     &                   &     \{2\} 1   &  1 & 1  & 0,0002    &        5 &              1  & A         \\
Hs.514777-S     &                   &     1         &  3 & 1  & 2,3E-52   &        7 &              1  & A         \\
\hline
GI\_18765712-S  &        \{7\}  1   &1&  \fbox{1} 0 & \fbox{1} 0& 1,2E-20 &\fbox{1} 3&   \fbox{1}  0   &  A/B           \\
GI\_22062109-S  &            {1}    &1&  \fbox{1} 0 & \fbox{1} 0& 2,6E-07 & \fbox{1} 2 & \fbox{1}  0   &  A/B          \\
GI\_42660576-S  &                   &     \{4\} 2   &  1 & 1   & 0,0002   &       2 &              1   &   A/B         \\
hmm25278-S      &                   &     \{3\} 2   &  4 & 1   & 9,1E-15  &       4 &              1   &   A/B         \\
hmm34610-S      &                   &           2   &  7 & 1   & 5,2E-07  &       4 &              1   &   A/B          \\
Hs.517172-S     & {\{11\} 2} &\{2\} 1 & \fbox{1} 2 & \fbox{1} 1 & 3,9E-05 &\fbox{2} 2 & \fbox{2}   1   &   A/B         \\
\hline
GI\_16753224-S  &                   &     1         &  1 &    & 4,9E-27   &       5  &                 &  B          \\
GI\_21237760-S  &                   &     1         &    &    & 3,2E-10   &       2  &                 &  B          \\
GI\_22325391-S  &                   &     \{2\} 1   &    &    & 2,4E-05   &       2  &                 &  B          \\
GI\_31543145-S  &                   &     1   & \fbox{1} 1 &  & 0,0015    & \fbox{1} 2  &              &  B          \\
GI\_34147394-S  &                   &     1         &    &    & 2,2E-06   &       1  &                 &  B          \\
GI\_37552433-S  &                   &    \{2\} 1    &  1 &    & 2,2E-18   &       3  &                 &  B      \\
GI\_38679899-S  &      {\{3\} 1 }   &     1         &    &    & 3,8E-06   &       2  &                 &  B          \\
GI\_38679979-A  &                   &     1         &    &    & 0,0003    &       0  &                 &  B          \\
GI\_40316914-S  &                   &     1         &    &    & 1,3E-08   &       3  &              1  &  B          \\
GI\_42659564-S  &                   &     \{4\} 1   &  2 &    & 1,9E-21   &       4  &                 &  B          \\
GI\_9790904-S   &                   &     1         &  1 &    & 1,5E-12   &       1  &                 &  B          \\
Hs.435267-S     &                   &     1         &    &    & 7,4E-12   &       2  &                 &  B          \\
\hline
GI\_10864076-S  &                   &   \{21\}     5&  1 & 1  & 1,5E-07   &        4  &             1  &  C           \\
GI\_19557676-S  & \fbox{98} 12 &  \{7\} 1 & \fbox{2} 1 & \fbox{2} 0  & 0,01 & \fbox{2} 1  & \fbox{2} 0   &  C           \\
GI\_23510353-S  &                   & \{27\}      5 &  1 & 1  &  0,24     &        2 &              1  &   C           \\
GI\_33469144-S  &                   &    \{57\}   1 &  1 &    &  0,0001   &        3 &                 &   C          \\
GI\_37537711-S  &                   &    \{17\} 5   &  3 & 1  &  1,6E-06  &        8 &              2  &   C           \\
GI\_41150880-S  &                   & \{42\}      11&  1 & 1  &  2,2E-15  &       10 &              3  &   C           \\
GI\_42657060-S  &                   & \{53\}      11&  3 & 4  &  1,0E-10  &        8 &              5  &   C

     \end{tabular}
\end{center}
\end{footnotesize}
\end{table}

We want to compare our model selection approach in particular with the analysis of \cite{Strange3}  for trans association of gene expression with SNPs. Pooling all four populations they found 44 genes showing trans association, where detailed results are provided in their Supplementary Table 6. 
To make our results comparable with \cite{Strange3}  we also restrict ourselves to additive models of the form (\ref{AddModel}), though we believe it would be interesting to consider additionally dominance effects. To account for population structure in our models we add dummy variables for populations CEU, CHB and JPT.

In \cite{Strange3} only some 25000 candidate SNPs were considered as putatively functional SNPs. Unfortunately these SNPs were not clearly specified (the corresponding link in the supplementary material is not providing the relevant list of SNPs), and we decided to search over all available SNPs. As a consequence mBIC2 will use a much larger penalty for multiple testing, on the other hand functional SNPs might be found which were not at all considered by \cite{Strange3}.

Starting point was the set of SNPs from phase 2 of the HapMap project. In accordance with \cite{Strange3} we only considered SNPs with MAF $> 0.05$, which results in a set of 2698476 SNPs. Filtering identical SNPs yields a subset of 2145627 SNPs, many of which are strongly correlated. In that situation many of the SNPs do not bring substantially new information to the genotype data. To solve this problem in \cite{BF} the notion of `effective number' of markers was introduced, an idea which can be also found e.g. in \cite{Nicod}. The `effective number' of markers is used to replace the number of available regressors in the penalty for modifications of BIC. In our real data analysis the `effective number' of SNPs is calculated based on the  clustering algorithm described in \cite{From}. This algorithm  yields clusters of SNPs which all have pairwise correlation above a chosen threshold $C$. In accordance with results of Section \ref{Sec:Sim} we have chosen $C = 0.7$, which leads to an effective number of 780675 SNPs. 

We performed model selection based on mBIC2 (\ref{mBIC2}) with  $p = 780675$ and we applied the search algorithm described in Section \ref{Sec:Search}.   We observed that in some cases the stepwise selection procedure got trapped in local minima for models which are too large. Therefore we added a final step to the search strategy, where we  performed an all subset selection over the combined set of SNPs detected by mBIC2 and by \cite{Strange3}.  If this set of SNPs was excessively large ($> 25$) we performed backward selection, and all subset selection only for models including less than 5 SNPs. 
Table \ref{Tab:Results_Real} shows a comprehensive summary of these results.

As discussed in Section \ref{Sec:threshold}  SNPs found by model selection are naturally representatives of a number of correlated SNPs.
On the other hand many  SNPs  detected by the multiple testing approach from \cite{Strange3}  are strongly correlated and frequently even have identical genotype for all individuals.  To make results comparable we selected representatives of correlated SNPs from \cite{Strange3} by applying Tagger  \cite{Bak}, a tag SNP selection algorithm   implemented in Haploview \cite{Bar}.  In accordance with the discussion in Section \ref{Sec:threshold} we used as threshold $|R| = 0.7$ (i.e. $R^2 = 0.49$).

Table \ref{Tab:Results_Real} provides  the number of tag SNPs for cis and for trans associations for the 44 genes with trans association reported in \cite{Strange3}. Furthermore we provide the number of cis and trans association detected when using mBIC2, first for models with dummy variables corresponding to different populations, then for models without such dummy variables. 
We also report the number of matches between \cite{Strange3} and our model selection approach, where we define that a match occurs when the absolute correlation between a tag SNP and a SNP detected by mBIC2 is larger than 0.7. 

For models considering population structure we report p-values of F Tests on the overall effect of the 3 dummy variables.
Taking into account population stratification is important. Without including dummy variables in most genes the number of detected trans SNPs is inflated. Almost all of these additional findings are associated with population structure, which corresponds well with the  small p-values observed in column three. For most genes the expression levels vary between populations, and among the huge number of SNPs there will always be some which pick up this variation.

Results are arranged according to the categories presented in the last column.
In category A we collect 19 genes where the model with dummy variables found all SNPs from \cite{Strange3},  in the 12 genes of category B it did not find any of them, and in the 6 genes of category A/B  it detected some but not all of them. Category C collects 7 genes for which \cite{Strange3} reports an extraordinary large number of  SNPs.
When we take into account population stratification, then for the majority of genes of category A and A/B our results are quite similar to those of \cite{Strange3}. In category B there are 7 genes for which Stranger reported 1 or 2 associated SNPs which were not detected using mBIC2. This is not surprising given the fact that we were penalizing for a much larger number of markers.

On the other hand, results for the genes hmm25278-S, hmm26651-S, hmm34610-S and hmm32074-S  are very interesting.  These genes are located on chromosome 1, 20, 8 and 6 respectively, but their expression levels are strongly correlated (pairwise correlation larger than 0.92 for each possible combinations).   \cite{Strange3} reported the following trans SNPs: rs9528181 for all four,  rs7318180 for the first three, and rs12860901 for the first two or them. These SNPs are all located on chromosome 13 at positions 113893447, 113835272 and 113901892, respectively, which indicates that this region on chromosome 13 has strong regulatory influence on the four genes under discussion. 

For all these genes model selection is finding larger models, which are summarized in Table \ref{Tab:Results_hmm}. 
All four models include a SNP on chromosome 13 which represents the detection of  \cite{Strange3}. Furthermore all four models include trans SNPs on chromosome 2 and on chromosome 3, though not all of them are located in proximity.  Apart from that there is a certain amount of ambiguity. For hmm25278-S there is one more SNP which does not correspond to any of the other detections. Models for the other three genes agree on SNP rs2044109, and they all include a SNP on chromosome 1.  The models of hmm32074-S and hmm34610-S include further non-corresponding SNPs. 
 In summary, 
according to our study there is  strong evidence that  more than one region has regulatory influence on the expression levels.  
Also this example shows that for the future a multivariate approach taking into account the information of correlated traits might be of some interest.

\begin{table}
\caption{Models selected for the genes hmm25278-S,  hmm26651-S, hmm32074-S and hmm34610-S.
SNP name (first line), chromosome and position (second line) for each selected SNP. 
 }\label{Tab:Results_hmm}

\begin{footnotesize}

\begin{center}
\begin{tabular}{rr|rr|rr|rr}
\multicolumn{2}{c|}{hmm25278-S}  & \multicolumn{2}{c|}{ hmm26651-S} & \multicolumn{2}{c|}{hmm32074-S} & \multicolumn{2}{c}{hmm34610-S}\\
\hline
\hline
\multicolumn{2}{c|}{rs9525262}   &          \multicolumn{2}{c|}{rs9525181}    & \multicolumn{2}{c|}{rs9525262} &            \multicolumn{2}{c}{rs9525262}                                          \\
13  &  113891161   &    13  & 113893447  &  13 &  113891161  &      13 &   113891161 \\
\hline
\multicolumn{2}{c|}{rs10048748}   &         \multicolumn{2}{c|}{rs10490450}   & \multicolumn{2}{c|}{rs17386102}  &            \multicolumn{2}{c}{rs10490450}                                         \\
2   &  165704573   &     2  &  33186442 & 2   & 17197272    &     2   &  33186442   \\
\hline
\multicolumn{2}{c|}{rs10937559}     &          \multicolumn{2}{c|}{rs6441934}  & \multicolumn{2}{c|}{rs1370718}  &           \multicolumn{2}{c}{rs6441934}                                      \\
3   &  194105335  &  3  &   45937806 &  3  &  32328135   &      3  &   45937806   \\
\hline
\multicolumn{2}{c|}{rs8028606}    &     \multicolumn{2}{c|}{rs2044109}    & \multicolumn{2}{c|}{rs2044109} &            \multicolumn{2}{c}{rs2044109}                                          \\
15    &  92857298  &  8  &   3074517 &   8  &  3074517    &      8  &   3074517   \\
\hline
&  &                  \multicolumn{2}{c|}{rs17455546} & \multicolumn{2}{c|}{rs2819755} &            \multicolumn{2}{c}{rs17455546}                                         \\
 &    &              1  &   100637742 & 1  &  236089656    &      1  &   100637742 \\ 
\hline
   &  &                 &  & \multicolumn{2}{c|}{rs13021147}  &           \multicolumn{2}{c}{rs3761945}                \\
   &  &                 &    &  2 &  107939438    &      1   &  228773391 \\
\hline
&   &     &       &  \multicolumn{2}{c|}{5 more SNPs on}   &           \multicolumn{2}{c}{rs17326215}               \\
  &     &     &     &   \multicolumn{2}{c|}{Chr. 5, 9, 10, 18, 22}  &      7  &   24408655  \\
     \end{tabular}
\end{center}
\end{footnotesize}
\end{table}


For the genes discussed above the three trans SNPs detected by \cite{Strange3} are located very close to each other on the same chromosome. This is actually typical:  For all 44 genes, the reported trans SNPs are located within a relatively small region. This holds  even for  the 7 genes of category C, which are characterized by an untypically large number of SNPs reported in \cite{Strange3}. These SNPs have a rather complex correlation structure, but their positions are for all cases within less than 400 kb.

 If we take for example gene  GI\_19557676-S, the reported cis SNPs (chromosome 6, between pos. 31105671 and 31439808) and trans SNPs (chromosome 6, between pos. 30045241 and 30049163) are located fairly close to each other. One might think of an extended cis region, and mBIC2 is finding 3 SNPs (2 cis, one trans) which represent the genetic variability within that region. Although the trans SNP rs3823342 (chr. 6,   pos. 30021046) found by model selection is not a match according to our definition based on correlation, it indicates the same region. The same is true for gene GI\_33469144-S, where SNP rs2996607    on chromosome 10 is in the same region as all the trans SNPs reported by \cite{Strange3}, though based on the correlation criterion it does not count as a match.

 If we look at the genes GI\_10864076-S (Chr. 16) and GI\_23510353-S (Chr. 19), in both cases mBIC2 found one matching trans SNP which turns out to be strongly correlated with all SNPs reported by \cite{Strange3}, namely $|R| > 0.49$ for GI\_10864076-S and  $|R| > 0.48$ for GI\_23510353-S. 
Now interestingly,  for genes GI\_37537711-S (Chr. 5), GI\_41150880-S (Chr. 18) and GI\_42657060-S (Chr. 4) the trans SNPs found by \cite{Strange3} are all lying exactly in the same region as those of GI\_10864076-S and GI\_23510353-S   (Chr. 6, between position 32500000 and 32800000), and also many trans SNPs are actually shared by these genes.  It is clear that this region must have a particularly  strong regulatory effect on other genes, that is susceptible to genetic variability. Multiple testing strategies pick up many correlated SNPs reflecting these signals, whereas mBIC2 is detecting a smaller number of SNPs representing that region.

Finally we want to mention several other genes for which additional trans SNPs have been found, namely GI 21536317-S, GI 31543145-S, GI 41147791-S, GI 42659564-S, GI 42662536-S,  Hs.514777-S and Hs.517172-S. Perhaps most remarkable among those are GI 41147791-S and GI 31543145-S, where the model selection approach was able to detect a cis SNP which was not detected by multiple testing.

\section{Discussion}

We have introduced a model selection approach for genome wide association studies using modifications of BIC which are based on sound theoretical considerations \cite{BZG, FBC}.  Elementary statistical arguments have shown that a model selection approach is preferable to multiple testing strategies, and a comprehensive simulation study confirmed this.  Finally we performed a real data analysis based on 210 individuals from the HapMap project, where model selection provided some interesting detections not found by the original analysis based on multiple testing.

Perhaps the most important result we obtained is that under complex models one cannot trust the order of p-values from single marker models. Test statistics are highly influenced by random small correlations to causal SNPs, leading on the one hand to a large number of false positives, on the other hand to severely reduced power. This loss of power might be one aspect of the
widely discussed phenomenon of missing heritability in GWAS (for a recent discussion see \cite{Man}).  It is believed that  missing heritability might be found in rare SNPs, or that epigenetic effects might play an important role \cite{McCH}. However, our results indicate that the statistical analysis performed is an important aspect of the problem, and that multiple testing strategies are just not really well suited for GWAS analysis.

In our simulation study model selection performed unambiguously better than multiple testing. In real data analysis, compared to the original analysis,  a substantial number of new putative regions of trans association could be found. Still, the effects were not as strong as in the simulation study; the largest model included 11 SNPs, two models were of size 7 and 5, the rest of size 4 or smaller. We believe that this is mainly due to the rather small sample size.  To select more complex models one would need studies with a larger number of individuals. Then it is expected that differences between multiple testing and model selection are getting even more pronounced.

To deal with the huge number of potential models we introduced a rather simple search strategy designed for this particular application. Our search strategy served well in the simulation study, but it had some limitations in the real data analysis. The focus of this manuscript was not on search strategies.  Modifications of ideas presented in \cite{Bai} in the context of QTL mapping might be useful. Other possible approaches have been discussed in \cite{Chen3, CC4}. The exact choice of model search strategies in GWAS is certainly a fruitful topic for further research.

{\bf Acknowlegments: }

This research was funded by the WWTF project MA09-007.





\bibliographystyle{elsarticle-num}


\begin{thebibliography}{00}

\bibitem[Abramovich {\em et~al.}(2006)]{ABDJ}
Abramovich F., Benjamini Y., Donoho D. L.,  Johnstone I. M.  Adapting to unknown sparsity by controlling the false discovery rate \textit{Ann. Statist.} 2006 \textbf{34}, 584-653.


\bibitem[\protect\citeauthoryear{Akaike}{Akaike}{1974}]{AIC}
Akaike, H. (1974)
\newblock {A new look at the statistical model identification.}
\newblock {\em IEEE Trans. Automat. Control\/}~{\em 19\/}, 716--723.


\bibitem[Baierl et al.(2006)]{Bai}
Baierl, A., Bogdan, M., Frommlet, F., Futschik, A. (2006) On Locating Multiple Interacting Quantitative Trait Loci in Intercross Designs. Genetics. {\bf 173}: 1693-1703.


\bibitem[\protect\citeauthoryear{Baierl, Futschik, Bogdan, and Biecek}{Baierl
  et~al.}{2007}]{BFBB}
Baierl, A., A.~Futschik, M.~Bogdan, and P.~Biecek (2007).
\newblock {Locating multiple interacting quantitative trait loci using robust
  model selection}.
\newblock {\em Computational Statistics and Data Analysis\/}~{\em 51},
  6423--6434.



\bibitem[Balding(2006)]{Bal}
Balding, D.~J., (2006)
A tutorial on statistical methods for
population association studies
Nat. Rev. Gen. {\bf 7}:781--791.



\bibitem[de Bakker et~al(2005)]{Bak} de Bakker, P.I., Yelensky, R., Pe'er, I., Gabriel, S.B., Daly, M.J., and Altshuler, D.   \ (2005)\ Efficiency and power in genetic association studies. {Nat Genet.}\ {\bf 37}, 1217--23.

\bibitem[Barrett et~al(2005)]{Bar} Barrett, J.C., Fry, B., Maller, J. and Daly, M.J., (2005) Haploview: analysis and visualization of LD and haplotype maps. {Bioinformatics. }\ {\bf 21}, 263--265.


\bibitem{BH}
\textsc{Benjamini, Y.} and \textsc{Hochberg, Y., } {(1995)}.
Controlling the false discovery rate: a practical and powerful
approach to multiple testing.
\textit{J. Roy. Statist. Soc. Ser. B.} \textbf {57} 289--300. MR{1325392}


\bibitem[Bogdan {\em et~al.}(2010)]{BCFG}
 Bogdan, M., Chakrabati, Frommlet, F., A. and Ghosh, J.K. (2010) Bayes oracle and the asymptotic optimality of the multiple testing procedures under sparsity. {\it To appear}, currently available at arXiv:1002.3501


\bibitem[Bogdan {\em et~al.}(2008)]{BCG}
Bogdan, M., Chakrabarti A., Ghosh, J.K. (2008). Optimal rules for multiple testing and sparse multiple regression, Technical Report I-18/08/P-003, Institute of Mathematics and Computer Science, Wroc{\l}aw University of Technology, 2008.



\bibitem[Bogdan et~al(2008)]{BF} Bogdan M.,  Frommlet F.,  Biecek P.,  Cheng R.,  Ghosh J.K.,  Doerge R.W.   \ (2008)\ Extending the Modified Bayesian Information Criterion (mBIC)
to Dense Markers and Multiple Interval Mapping. {Biometrics}\ {\bf 64}, 1162--1169.



\bibitem[Bogdan {\em et~al.}(2004)]{BGD}
Bogdan, M., Ghosh, J.~K., and Doerge, R.~W. (2004).
\newblock Modifying the schwarz bayesian information criterion to locate
  multiple interacting quantitive trait loci.
\newblock {\em Genetics}, {\bf 167}, 989--999.


\bibitem[Bogdan {\em et~al.}(2007)]{BGOT}
Bogdan, M., Ghosh, J.~K., Ochman, A. and  Tokdar S.T. (2007)
\newblock On the Empirical Bayes approach to the problem of multiple
testing.  \newblock {\em Quality and Reliability Engineering International}, {\bf 23}, 727--739.

\bibitem[Bogdan {\em et~al.}(2008)]{BGT}
Bogdan, M., Ghosh, J. K. and Tokdar S. T. (2008)
 \newblock A comparison of the Simes-Benjamini-Hochberg procedure with some Bayesian
 rules for multiple testing.
 \newblock IMS Collections, {\bf Vol.1}, Beyond Parametrics in Interdisciplinary
 Research: Fetschrift in Honor of Professor Pranab K. Sen, edited by N. Balakrishnan, Edsel Pe\~na and
 Mervyn J. Silvapulle, pp. 211--230, Beachwood Ohio.

 \bibitem[Bogdan {\em et~al.}(2008)]{BZG}
 Bogdan, M.,  \.Zak-Szatkowska, M., Ghosh, J.K. Selecting explanatory variables with the modified version of Bayesian Information Criterion, {\it Quality and Reliability Engineering International},
{\bf 24}: 627--641,  2008.


\bibitem[Broman and Speed (2002)]{BS}
Broman, K.W., Speed, T.P. (2002)
A model selection approach for the identification of quantitative trait loci in experimental crosses,
\textit{Journal of the Royal Statistical Society: Series B (Statistical Methodology)} {\bf 64} (4): 641--656.


\bibitem[\protect\citeauthoryear{Chen and Chen}{Chen and Chen}{2008}]{Chen1}
Chen, J. and Z.~Chen (2008).
\newblock {Extended Bayesian Information criteria for model selection with
  large model spaces}.
\newblock {\em Biometrika\/}~{\em 95\/}(3), 759--771.

\bibitem[\protect\citeauthoryear{Chen and Chen}{Chen and
  Chen}{2010}]{Chen2}
Chen, J. and Z.~Chen (2010).
\newblock {Extended BIC for small $n$-large-$P$ sparse GLM}.
\newblock submitted, available at  {\textit{www.stat.nus.edu.sg/{$\sim$}stachenz}/ChenChen.pdf}.

\bibitem[Chen and Chen (2010b)]{CC4}
Chen, J. and Z.~Chen (2010).
Tournament screening cum EBIC for feature selection with high-dimensional feature spaces
\textit{Science in China Series A: Mathematics} {\bf 52} (6): 1327--1341.



\bibitem[Chipman et al. (2001)]{CGM}
Chipman, H., George, E.I. and McCulloch, R.E. (2001)
The practical implementation of
Bayesian model selection (with discussion). In \textit{Model Selection} (P. Lahiri, ed.)
 66--134.
IMS, Beachwood, OH.

\bibitem[Cho   et al. (2009)]{Cho}
Cho Y.S., et al. (2009)
A large-scale genome-wide association study of Asian populations uncovers genetic factors influencing eight quantitative traits.
\textit{Nat Genet.} 41(5):527--534.

\bibitem[Colditz and  Hankinson (2005)]{Col}
Colditz G.A. and Hankinson S.E. (2005)
The Nurses' Health Study: lifestyle and health among women.
\textit{Nature Reviews Cancer} 5, 388--396


\bibitem[Dudbridge and Gusnanto (2008)]{DG}
Dudbridge, F., Gusnanto, A. (2008)
Estimation of Significance Thresholds for Genomewide
Association Scans  \textit{Genet. Epid.} {\bf 32}: 227--234

\bibitem[\protect\citeauthoryear{Erhardt, Bogdan and Czado}{Erhardt  et~al.}{2010}]{EBC}
Erhardt, V., M. Bogdan and C. Czado (2010).
 \newblock{Locating multiple interacting quantitative trait loci with the zero-inflated generalized Poisson regression,} {\it Statistical Applications in Genetics and Molecular Biology},  {\bf Vol 9 : Iss. 1}, Article 26.



\bibitem[Frommlet (2010)]{From}
Frommlet, F. (2010). Tag SNP selection based on clustering according to dominant sets found using replicator dynamics,  \textit{Adv Data Anal Classif} \textbf{4}: 65--83

\bibitem[Frommlet et al. (2010)]{FBC}
Frommlet, F., {Bogdan, M.} and {Chakrabarti, A.} (2010). Asymptotic Bayes optimality under sparsity of selection rules for general priors.
\textit{Technical report}, arXiv:1005.4753


\bibitem[George and Foster (2000)]{GF}
{George, E.I.} and {Foster, D.P.} (2000). Calibration and empirical Bayes variable selection.
\textit{Biometrika.} \textbf{87}: 731--747.


\bibitem[Ganesh et al. (2009)]{Gan}
Ganesh S.K., et al. (2009)
Multiple loci influence erythrocyte phenotypes in the CHARGE Consortium.
\textit{Nat Genet.} {\bf 41(11)} 1191--1198.


\bibitem[HapMap Consortium(2007)]{Hap}The International HapMap Consortium. \ (2007) \ A second generation human haplotype map of over 3.1 million SNPs. Nature {\bf 449}, 851--862.


\bibitem[Hirschhorn  and Daly (2005)]{HD}
Hirschhorn, J.N. and Daly, M.J. (2005) Genome-wide association studies for common diseases and complex traits. Nat Rev Genet.  {\bf 6}(2):95--108.

\bibitem[Koren  et al. (2006)]{Kor}
Koren M, Kimmel G, Ben-Asher E, Gal I, Papa MZ, Beckmann JS, Lancet D, Shamir R, Friedman E.  (2006).
ATM haplotypes and breast cancer risk in Jewish high-risk women.
\textit{Br J Cancer.} {\bf 94}(10): 1537--1543.


\bibitem[Lao et al. (2008)]{Lao}
Lao O., et al.
(2008)
Genome-wide association studies for complex traits: consensus, uncertainty and challenges.
  \textit{Curr Biol.} {\bf 18}(16):1241--1248.


\bibitem[Madow (1940)]{Mad}
Madow, W. (1940) The distribution of quadratic forms in noncentral normal random variables. \textit{Ann. Math. Stat.} {\bf 11}, 100--103).


\bibitem[Manolio et al. (2009)]{Man}
Manolio TA, Collins FS, Cox NJ, Goldstein DB, Hindorff LA, Hunter DJ, McCarthy MI, Ramos EM, Cardon LR, Chakravarti A, Cho JH, Guttmacher AE, Kong A, Kruglyak L, Mardis E, Rotimi CN, Slatkin M, Valle D, Whittemore AS, Boehnke M, Clark AG, Eichler EE, Gibson G, Haines JL, Mackay TF, McCarroll SA, Visscher PM. (2009) Finding the missing heritability of complex diseases.
 \textit{Nature.} {\bf 461}(7265):747--753.


\bibitem[McCarthy et al. (2008)]{McC}
McCarthy MI, Abecasis GR, Cardon LR, Goldstein DB, Little J, Ioannidis JP, Hirschhorn JN. (2008)
Genome-wide association studies for complex traits: consensus, uncertainty and challenges.
  \textit{Nat Rev Genet.} {\bf 9}(5):356--369.

\bibitem[McCarthy and Hirschhorn (2008)]{McCH}
McCarthy, M.I. and Hirschhorn, J.N. (2008) Genome-wide association studies: potential
next steps on a genetic journey. \textit{Hum. Mol. Genet.} {\bf 17}, R156--R165).


\bibitem[Meisinger et al. (2009)]{Mei}
Meisinger C.,  et al. (2009)
A genome-wide association study identifies three loci associated with mean platelet volume.
\textit{Am J Hum Genet.} {\bf 84} 66--71.


\bibitem[Nelson et al. (2008)]{Nel}
Nelson M.R.,  et al. (2008)
\textit{Am J Hum Genet.} {\bf 83} 347--58. Epub 2008 Aug 28.


\bibitem[Newton-Cheh  et al. (2009)]{New}
Newton-Cheh C, et al. (2009)
Genome-wide association study identifies eight loci associated with blood pressure.
\textit{Nat Genet.} [Epub ahead of print]


\bibitem[Nicodemus et~al(2005)]{Nicod}  Nicodemus K.K, Liu W, Chase G.A., Tsai Y.Y., Fallin M.D. \ (2005)\ Comparison of type I error for multiple test corrections in large single-nucleotide polymorphism studies using principal components versus haplotype blocking algorithms.
{BMC Genet.}\ {\bf 6}(Suppl 1).


\bibitem[Ouwehand (2009)]{Ouw}
Ouwehand W.H. (2009)
The discovery of genes implicated in myocardial infarction.
\textit{J Thromb Haemost} {\bf 7} Suppl 1:305--307.


\bibitem[Potkin et al. (2009a)]{Pot_a}
Potkin S.G., Guffanti G., Lakatos A., Turner J.A., Kruggel F., Fallon J.H., Saykin A.J., Orro A., Lupoli S., Salvi E., Weiner M., Macciardi F. (2009)
Hippocampal atrophy as a quantitative trait in a genome-wide association study identifying novel susceptibility genes for Alzheimer's disease.
\textit{PLoS One} {\bf 4(8)}: e6501


\bibitem[Potkin et al. (2009b)]{Pot_b}
Potkin,S.G., Turner, J.A., Guffanti,G.,
Lakatos, A., Torri, F., Keator,  D.B., and
Macciardi, F. (2009)
Genome-wide strategies for discovering genetic
influences on cognition and cognitive disorders:
Methodological considerations,
\textit{Cognitive Neuropsychiatry}  {\bf 14}: (4/5), 391--418

\bibitem[\protect\citeauthoryear{Schwarz}{Schwarz}{1978}]{SCH}
Schwarz, G. (1978)
\newblock{ Estimating the dimension of a model.}
\newblock {\em Annals of Statistics\/}~{\em 6\/}, 461--464.

\bibitem[Stranger et al. (2005)]{Strange1}
Stranger, B.E, M.S. Forrest, A.G. Clark, M.J. Minichiello, S. Deutsch, R. Lyle, S. Hunt, B. Kahl, S.E. Antonarakis, S. Tavare, P. Deloukas, E.T. Dermitzakis. (2005). Genome-wide associations of gene expression variation in humans. \textit{PLoS Genetics} {\bf1}:e78.



\bibitem[Stranger et al. (2007a)]{Strange2}
Stranger, B.E., M.S. Forrest, M. Dunning, C.E. Ingle, C. Beazley, R. Redon, C.P. Bird, A. de Grassi, C. Lee, C. Tyler-Smith, N. Carter, S.W. Scherer, S. Tavar�, P. Deloukas, M.E. Hurles, E.T. Dermitzakis. (2007) Relative impact of nucleotide and copy number variation on gene expression phenotypes. \textit{Science} {\bf 315}: 848--853.


\bibitem[Stranger et al. (2007b)]{Strange3}
Stranger, B.E., A.C. Nica, M.S. Forrest, A. Dimas, C.P. Bird, C. Beazley, C.E. Ingle, M. Dunning, P. Flicek, S. Montgomery, S. Tavar�, P. Deloukas, E.T. Dermitzakis. (2007). Population genomics of human gene expression. \textit{Nature Genetics} {\bf  39}: 1217--1224.


\bibitem[Wei et al. (2009)]{Wei}
Wei, Z., Sun, W., Wang, K. and Hakonarson, H. (2009)
Multiple testing in genome-wide association studies via hidden Markov models,
\textit{Bioinformatics}  {\bf 25}:(21), 2802--2808


\bibitem[Ziegler et al. (2008)]{ZKT}
Ziegler, A., K\"onig, I. R., and Thompson, J.R. (2008)
Biostatistical Aspects of Genome-Wide Association Studies,
\textit{Biometrical Journal}  {\bf 50}:1, 8--28

\bibitem[Zhang et al. (20068)]{Zha}
Zhang, C.L., Qi, D.J., Hunter, J.B., Meigs, J.E., Manson, J.E., vna Dam, R.M., Hu, F.B. (2006)
Variant of transcription factor 7-like 2 (TCF7L2) gene and the risk of type 2 diabetes in large cohorts of U.S. women and men. \textit{Diabetes} {\bf 55}: 2645--2648.

\bibitem[\protect\citeauthoryear{\.Zak, Baierl, Bogdan, and Futschik}
{\.Zak et~al.}{2007}]{ZBBF}
\.Zak, M., A.~Baierl, M.~Bogdan, and A.~Futschik (2007).
\newblock {Locating multiple interacting quantitative trait loci using
  rank-based model selection}.
\newblock {\em Genetics\/}~{\em 176\/}(3), 1845--1854.

\bibitem[\protect\citeauthoryear{\.Zak-Szatkowska and Bogdan}{\.Zak-Szatkowska and Bogdan}{2010}]{Zak:2010}
\.Zak-Szatkowska M. and M. Bogdan (2010).
\newblock{Applying generalized linear models for identifying important factors in large data bases.}
\newblock{{\em {Technical Report I-18/2010/P-001}\/}}.
\newblock Institute of Mathematics and Computer Science, Wroclaw University of
  Technology, \textit{www.im.pwr.wroc.pl/{$\sim$}mbogdan/Preprints}.


\bibitem[\protect\citeauthoryear{Zhao and Chen}{Zhao and
  Chen}{2010}]{Chen3}
Zhao, J. and Z.~Chen (2010).
\newblock {A two-stage penalized logistic regression approach to case-control genome-wide association studies.}
\newblock submitted, available at  { \textit{www.stat.nus.edu.sg/{$\sim$}stachenz/MS091221PR.pdf}}.




\end{thebibliography}







\end{document}